\documentclass{article}
\usepackage{comment}

\PassOptionsToPackage{numbers,sort&compress}{natbib}
\usepackage[preprint]{neurips_2026}

\usepackage[utf8]{inputenc}
\usepackage[T1]{fontenc}
\usepackage{hyperref}
\usepackage{url}
\usepackage{booktabs}
\usepackage{amsfonts}
\usepackage{amsmath}
\usepackage{amssymb}
\usepackage{nicefrac}
\usepackage{microtype}
\usepackage[dvipsnames]{xcolor}
\usepackage{graphicx}
\usepackage{algorithm}
\usepackage{algorithmic}
\usepackage{multirow}
\usepackage{subcaption}
\usepackage{tikz}
\usepackage{pgfplots}
\pgfplotsset{compat=1.18}
\usepackage{subcaption}
\usepackage{float}
\usepackage{placeins}
\usepackage{cleveref}
\usepackage{float}
\usetikzlibrary{arrows.meta, positioning, shapes.geometric, fit, calc}

\usepackage{tcolorbox}
\definecolor{takeawaytint}{HTML}{EDF4F3}
\newtcolorbox{takeawaybox}{
  colback=takeawaytint,
  colframe=takeawaytint,
  boxrule=0pt,
  arc=1.5mm,
  boxsep=2pt,
  left=7pt, right=7pt, top=4pt, bottom=4pt
}

\newif\ifcomments
\commentsfalse
\ifcomments
    \providecommand{\shubham}[1]{{\color{blue}{/* shubham: #1 */}}}
    \providecommand{\mert}[1]{{\color{olive}{/* mert: #1 */}}}
    \providecommand{\andrei}[1]{{\color{teal}{/* andrei: #1 */}}}
    \providecommand{\shu}[1]{{\color{magenta}{/* shu: #1 */}}}
    \providecommand{\ion}[1]{{\color{cyan}{/* ion: #1 */}}}
    \providecommand{\ak}[1]{{\color{orange}{/* alex: #1 */}}}
    \providecommand{\melissa}[1]{{\color{OrangeRed}{/* melissa: #1 */}}}
    \providecommand{\datacheck}[1]{{\color{blue}{/* data check: #1 */}}}
\else
    \providecommand{\shubham}[1]{}
     \providecommand{\andrei}[1]{}
    \providecommand{\mert}[1]{}
    \providecommand{\shu}[1]{}
    \providecommand{\ion}[1]{}
    \providecommand{\ak}[1]{}
    \providecommand{\melissa}[1]{}
    \providecommand{\datacheck}[1]{}
\fi

\newcommand{\methodname}{\textbf{AdaMAST}}
\newcommand{\fullname}{Adaptive Multi-Agent System Failure Taxonomies}

\definecolor{adamastpurple}{HTML}{4C1D95}
\definecolor{adamastcoral}{HTML}{7F1D1D}
\definecolor{adamastgreen}{HTML}{064E3B}

\title{Fantastic Adaptive Taxonomies and How to Use Them}

\author{%
  \mdseries 
  Mert Cemri$^{1*}$ \quad
  Andrei Cojocaru$^{1*}$ \quad
  Melissa Pan$^{1}$ \quad
  Shu Liu$^{1}$ \quad
  Shubham Agarwal$^{1}$ \\
  Alexander Krentsel$^{1}$ \quad
  Jay Tang$^{2}$ \quad
  Kannan Ramchandran$^{1}$ \quad
  Joseph E.~Gonzalez$^{1}$ \\
  Matei Zaharia$^{1}$ \quad
  Alex Dimakis$^{1,3}$ \quad
  Ion Stoica$^{1}$ \\[6pt]
  $^{1}$University of California, Berkeley \quad
  $^{2}$Apple \quad
  $^{3}$Bespoke Labs
}

\begin{document}

\maketitle

\begingroup
\renewcommand\thefootnote{}%
\footnotetext{$^{*}$Equal contribution. Correspondence: \texttt{cemri@berkeley.edu}. \\ Code: \url{https://github.com/multi-agent-systems-failure-taxonomy/AdaMAST}.}%
\addtocounter{footnote}{-1}%
\endgroup

\begin{abstract}
An agent system's execution traces record how it fails, and procedures that improve such a system without changing model weights (trajectory selection, prompt and workflow optimization, runtime monitoring) read these traces for feedback. Yet raw traces are a poor medium for accumulating such feedback: they are long, instance-specific, and provide no stable vocabulary for recurring failures. We argue that an agent system should instead maintain an explicit representation of how it fails, induced from its own behavior and reusable wherever failure feedback is needed. \methodname{} builds this representation by converting a target system's traces into a compact, evidence-grounded failure taxonomy. The taxonomy consists of named failure codes organized along three fixed axes---system-level, role-specific, and domain-specific---while every code name, definition, and evidence pattern is induced from the traces. No code is hand-authored, and no individual trace requires human annotation. The resulting taxonomy is not merely a post-hoc diagnostic, but a shared feedback interface. 
We demonstrate \methodname{} improves agents in three ways. In agent-system search, taxonomy-coded diagnoses of failed candidates outperform free-form reflection on all five benchmarks we test. At runtime, taxonomy feedback raises SWE-agent's resolution on SWE-bench Verified Mini from $60\%$ with free-text reflection to $70\%$, and improves Claude Code from $64.0\%$ to $70.7\%$ when deployed as a runtime skill. In trajectory selection, \methodname{}-Judge, a verifier built on the induced codes, improves best-of-$5$ accuracy on Terminal-Bench~2.0 by $8$--$15$ points over Pass@1. Beyond these downstream gains, the induced vocabulary is \emph{compact}, compressing the failure-relevant content of traces by an order of magnitude while largely preserving their distinctions; \emph{human-faithful}, matching expert failure annotations more closely than a hand-crafted reference vocabulary; and \emph{adaptive}, with taxonomies induced for different domains sharing only a small fraction of their codes. Adaptive failure taxonomies close the loop between the traces agents produce and the procedures that improve them.
\end{abstract}

\section{Introduction}
\label{sec:intro}
Most procedures that improve an LLM agent system without changing model weights consume its execution traces: verifiers ranking best-of-$N$ attempts, optimizers rewriting the agent after failed runs, monitors inspecting an ongoing run before an action commits. The feedback these procedures extract from a trace takes one of two forms today, and each loses something essential. A scalar outcome (a pass rate, a verifier score) persists and compares cleanly across runs, but records only that a trajectory failed, never why. A free-text critique \citep{shinn2023reflexion,madaan2023selfrefine} does name the why, but is derived from scratch and discarded with the run: the recurring failures of that particular system (a checker that rubber-stamps its solver, a tool call retried unchanged) are re-diagnosed on every run, in prose no other consumer can reuse. Our central claim is that the raw trace need not double as the feedback interface: what can serve instead is an artifact durable like a score and diagnostic like a critique, a compact vocabulary of named failure modes, produced once from the system's own traces. Each consumer still reads the run it acts on; what it no longer needs is the accumulated trace pool, which the vocabulary can replace at a fraction of the context cost (\Cref{sec:compression}).

What should this vocabulary contain? MAST \citep{cemri2025mast}, the reference catalogue for multi-agent failures, distills fourteen recurring failure modes from hundreds of annotated traces, and it remains a strong baseline throughout our experiments.  But a catalogue fixed before the target system is observed faces a limit that is structural, not a consequence of quality: role-specific failures presuppose roles that exist only once an architecture is instantiated, and domain-specific failures presuppose task knowledge no general catalogue enumerates. No pre-committed list anticipates a competitive-programming agent committing to a greedy strategy where the problem requires dynamic programming, or a STEM agent producing a numerically coherent derivation that violates a physical law. The gap is measurable: taxonomies we induce for six domains share codes at a mean pairwise Jaccard of only $0.14$ (\Cref{sec:taxonomy_variation}); were a fixed checklist sufficient, induction would keep rediscovering it. The diagnostic vocabulary for a specific agent system must be induced from that system's own traces: nothing else has seen its tools, its roles, or its domain.

We present \methodname{} (\fullname{}),\footnote{The name follows our earlier work MAST~\citep{cemri2025mast}. The \emph{multi-agent} in the name is lineage rather than scope: the axes apply to any agentic system, and several of the systems we improve in this paper are single-agent, where the role axis names execution phases or remains empty.} which induces exactly this vocabulary. From a target system's traces, it produces a compact set of named failure codes organized across three fixed axes: system-level, role-specific, and domain-specific (\Cref{fig:failure_categories}). Only the axes are fixed, so that taxonomies stay comparable across systems and every code maps to an intervention point. The codes themselves (names, definitions, role labels, and evidence patterns) are induced from the traces: no code is hand-authored, no trace human-annotated.  A taxonomy is accepted only when independent annotators apply it consistently to held-out traces, so the artifact is auditable before anything consumes it (\Cref{sec:taxonomy}). Moreover, as the system evolves, online refinement merges, adds, or relabels codes, so the vocabulary tracks the \emph{current} system, rather than a snapshot of its past (\Cref{sec:mutation}).

The induced taxonomy is a feedback interface, and we evaluate it as one: first the artifact itself, then the procedures that consume it. The artifact is human-faithful, a property LLM--LLM agreement cannot certify: under a matched annotation panel, the induced vocabulary recovers expert failure labels on TRAIL better than the benchmark's own hand-crafted vocabulary (\Cref{sec:taxonomy_analysis}). The claim is amortization, not prompt formatting (\cref{app:feedback_form_diagnostic}): each target system's taxonomy is induced and validated once, then consumed unchanged by every procedure that reads its traces.

\paragraph{Contributions.}
\begin{itemize}
    \item \textbf{Adaptive failure taxonomies from traces.} A pipeline that induces compact, evidence-grounded failure taxonomies from a target system's execution traces along three fixed axes (system-level, role-specific, domain-specific), with no hand-authored codes and no per-trace human annotation, gated by inter-annotator agreement before deployment and refined online as the system evolves (\Cref{sec:taxonomy}).

    \item \textbf{The taxonomy as a measurable feedback interface.} Independent of downstream gains, the induced vocabulary is compact without collapsing distinctions (${\sim}18\times$ compression, $89\%$ unique code signatures; \Cref{sec:compression}), aligns with expert annotations better than a hand-crafted vocabulary under a matched panel ($\kappa=0.682$ vs.\ $0.516$ on TRAIL; \Cref{sec:trail_main}), and genuinely adapts to its target (mean cross-domain Jaccard $0.14$; \Cref{sec:taxonomy_variation}).

    \item \textbf{One interface, three consumers.} Taxonomies induced by the same pipeline, consumed unchanged within each target system, improve agent-system search on five benchmarks (\Cref{sec:evolution_results}), raise SWE-agent's resolution on SWE-bench Verified Mini from $60\%$ (Reflexion) and $68\%$ (MAST) to $70\%$ via an in-context integration (\Cref{sec:runtime_results}), and improve best-of-$5$ selection on Terminal-Bench~2.0 by $8$--$15$ points over Pass@1, retaining a $+3.4$ to $+4.5$ point margin over the fixed MAST vocabulary on the two non-saturated harnesses (\Cref{sec:terminalbench}).
\end{itemize}

\section{Related Work}
\label{sec:related}

\paragraph{Failure taxonomies for agent systems.}
A growing line of work shows that categorical failure analysis makes LLM agents easier to inspect, compare, and debug than scalar success rates alone. MAST~\citep{cemri2025mast} introduced an empirically grounded taxonomy of 14 multi-agent failure modes organized into system-design, inter-agent, and verification failures; TRAIL~\citep{deshpande2025trail} provides expert-annotated GAIA traces under a hand-crafted error taxonomy for agent debugging; and reliability failures have been characterized in production deployments~\citep{pan2026measuringagentsproduction}. Agent GPA~\citep{jia2025gpa} moves toward adaptivity by generating domain-specific evaluation criteria with LLM judges under a fixed goal--plan--action rubric, but the criteria serve post-hoc evaluation rather than a persistent vocabulary consumed by improvement procedures. Failure attribution~\citep{zhang2025whowhen} asks the complementary question of \emph{which agent and which step} caused a failure, and finds that even strong LLMs struggle to answer it from raw logs, underscoring the need for structured failure evidence. \methodname{} builds on this diagnostic view but changes both the \emph{source} and the \emph{role} of the taxonomy. The source: instead of fixing the vocabulary before the agent is observed, \methodname{} induces it from the target system's own traces, and the difference is measurable rather than cosmetic. Under a matched annotation panel and prompts, induced codes align with expert failure-area labels better than a hand-crafted reference vocabulary (\Cref{sec:trail_main}), and vocabularies induced for six domains share only $14\%$ of their codes (\Cref{sec:taxonomy_variation}). 
The role: the induced codes serve as a feedback interface consumed by improvement procedures, rather than as post-hoc analysis. Fixed catalogues remain strong baselines (MAST reaches parity with \methodname{} in some runtime settings; \Cref{sec:runtime_results}), which sharpens rather than undercuts the distinction: what an adaptive vocabulary adds is the role- and domain-specific long tail that no pre-committed catalogue can anticipate.

\paragraph{Improving agents from execution traces.}
Many methods improve LLM agents by reading traces and producing feedback, without updating model weights. Reflexion~\citep{shinn2023reflexion} and Self-Refine~\citep{madaan2023selfrefine} generate free-text critiques of failed attempts; LLM-as-a-judge and trace-verification methods select among candidate trajectories or score intermediate reasoning and tool-use steps~\citep{zheng2023llmjudge,kwok2026llmverifier,lightman2024prm}; and structured-feedback methods compute textual gradients or execution-graph feedback~\citep{yuksekgonul2024textgrad,cheng2024trace}, train critics or generative verifiers~\citep{mcaleese2024criticgpt,zhang2024generativeverifier}, or discover behavioral tests and error slices~\citep{ribeiro2020checklist,eyuboglu2022domino}. In all of these, the diagnosis is computed per trace or per optimization step, consumed by one procedure, and then discarded; the next run re-derives it. \methodname{} differs in the object it produces, not merely in how feedback is formatted: it amortizes trace analysis into a persistent, named, human-validated vocabulary that is induced once and then reused across runs and across procedures. The same codes serve as judging criteria for selection, diagnoses for search, and checkpoint guidance at runtime.

\paragraph{Search-based optimization of agent systems.}
Recent work optimizes LLM systems by searching over prompts, workflows, agent roles, or executable programs. Prompt- and workflow-level methods include Promptbreeder~\citep{fernando2023promptbreeder}, EvoPrompt~\citep{guo2024evoprompt}, DSPy~\citep{khattab2024dspy}, GEPA~\citep{agrawal2025gepa}, and AFlow~\citep{aflow2025}; program- and system-level methods include FunSearch~\citep{romeraparedes2024funsearch}, AlphaEvolve~\citep{novikov2025alphaevolve}, AdaEvolve~\citep{adaevolve2025}, and EvoX~\citep{liu2026evox}; and multi-agent design methods such as ADAS~\citep{hu2025adas}, EvoAgent~\citep{yuan2024evoagent}, and GPTSwarm~\citep{zhuge2024gptswarm} search over agent topologies or collaboration patterns. These searchers consume scalar scores or free-form reflections as their feedback signal. \methodname{} is orthogonal to them: it prescribes no search algorithm, mutation operator, or workflow representation; it supplies the diagnostic signal such a search consumes. Rather than proposing a competing optimizer, we plug \methodname{} into an existing one (AdaEvolve); our search results are therefore evidence about the vocabulary, not about a new search method.

\section{Method}
\label{sec:method}

We first define the terminology. A \textbf{failure mode} is a recurring way an agent
execution deviates from its intended outcome (e.g.,
\texttt{Premature\_Reasoning\_Truncation}: ``agent stops a multi-step chain
before reaching a final answer''), characterized by a name, description, and
evidence pattern. A \textbf{failure taxonomy} is a structured catalogue of such
modes; each entry is a \textbf{failure code}. The taxonomy is the artifact
this paper produces. It is inert on its own: it is exercised by
\textbf{consumers}, trace-reading procedures that would otherwise diagnose raw
trajectories from scratch in free text. We call a consumer
\textbf{taxonomy-conditioned} when the failure-relevant portion of its output
is constrained to the taxonomy's code set: it reports which codes fired, on
which evidence, and with what supporting excerpts. Conditioning is a property
of this interface, not of any particular component. The conditioned reader may be a separate LLM call that inspects trace evidence and returns a verdict (a \textbf{judge}), the acting agent itself checking its own trace mid-run, or a bank of per-code scoring prompts inside a verifier. Judges predate this work and require no taxonomy: an unconditioned judge simply returns whatever form its prompt requests: a scalar score, a free-text critique, or an unconstrained structured diagnosis.
\methodname{} introduces none of these consumers; it supplies the vocabulary
they share.

\begin{figure}[!h]
\centering
\includegraphics[width=\linewidth]{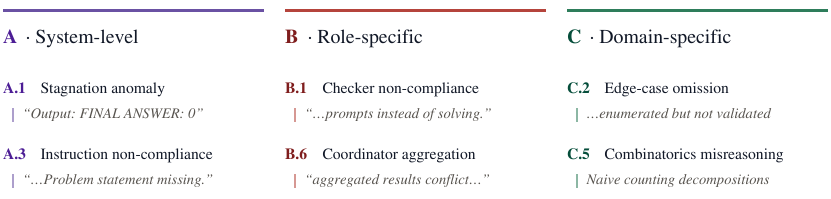}
\caption{\textbf{Three adaptive failure axes.} The axes of intervention are
fixed; the concrete code names, definitions, role labels, and evidence
patterns are induced from the target system's traces. Shown: representative
codes from the OlympiadBench evolution run, with verbatim evidence excerpts
from the run's judge outputs in quoted italics
(Appendix~\ref{app:breakthrough_judge}); unquoted lines under C-codes are
induced code descriptions (Table~\ref{tab:olympiad_taxonomy_top5}).}
\label{fig:failure_categories}
\end{figure}

Given execution traces from a target system, \methodname{} \emph{induces} a
compact, target-specific taxonomy organized along the three fixed axes introduced in
\Cref{sec:intro}: \textbf{Axis~A} (system-level), \textbf{Axis~B}
(role-specific), and \textbf{Axis~C} (domain-specific). By \emph{induce} we
mean inductive generalization from a sample of traces to a reusable
vocabulary: every code is generated from, and grounded in, observed trace
evidence rather than authored in advance. We avoid \emph{infer} deliberately, since it would
wrongly suggest a hidden ground-truth taxonomy waiting to be recovered, when there is none. The axes partition
failures by intervention point: a failure belongs to Axis~A if the remedy is
to repair the harness or orchestration around the agents, to Axis~B if it is
to rewire a discovered role, and to Axis~C if it is to inject task knowledge. Throughout the paper we refer to
individual codes by axis letter and index (e.g., A.3, B.6, C.2).
\Cref{fig:failure_categories} illustrates the axes; a full generated taxonomy
for Frontier-CS, including firing frequencies, appears in
Appendix~\ref{app:taxonomy_examples}.

We treat the three axes as a practical organizational scaffold rather
than a uniquely optimal partition. In a three-seed granularity ablation,
the full taxonomy is directionally higher than a flat code list by
3.3 percentage points on TheoremQA and 1.7 percentage points on
MMLU-Pro, but neither difference is statistically conclusive
(\Cref{app:extended_design_ablations}).

The axes differ in how candidate codes are sourced, and hence in when codes
can appear at all. System-level and domain-specific candidates are seeded by
broad priors that the pipeline's own \textsc{analysis} phase proposes (coarse
architectural-risk patterns and common domain error modes, suggested before
any individual trace is examined for supporting evidence; no prior is authored
offline by a human) and a seeded candidate is retained only when trace
evidence supports it. Role-specific candidates are fully trace-induced: the
relevant roles are themselves discovered from the target architecture and
admit no general catalogue. Role codes therefore exist only when the
architecture contains differentiated roles: the multi-agent Frontier-CS
pipeline induces 13 role codes, while a flat solver--verifier pipeline on
TheoremQA induces none.

\subsection{Generating the taxonomy}
\label{sec:taxonomy}

\begin{figure}[h]
    \centering
    \includegraphics[width=\linewidth]{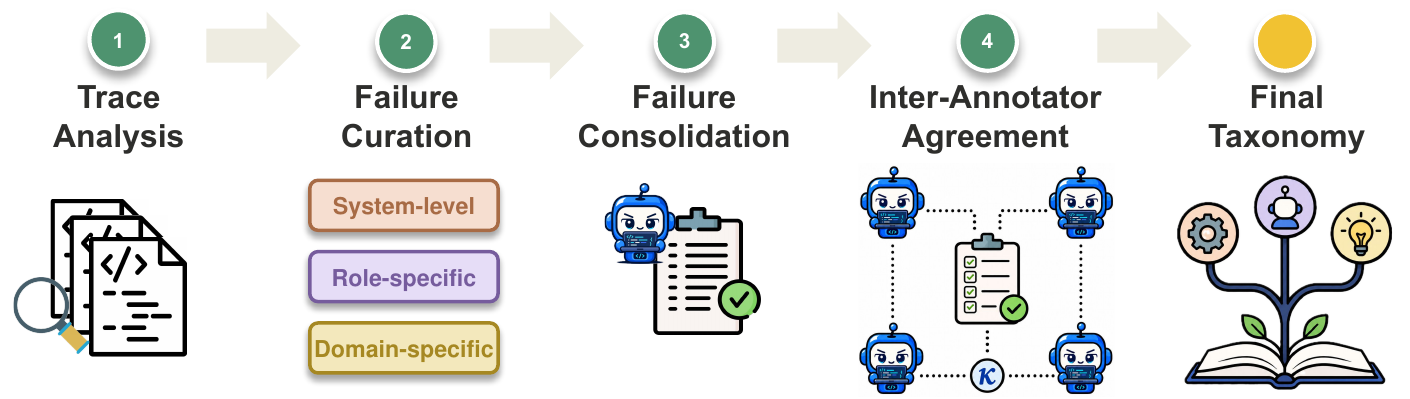}
    \caption{\textbf{\methodname{} pipeline.} \textsc{analysis},
\textsc{curation}, and \textsc{consolidation} produce a draft taxonomy
through an eight-step induction pipeline
(Appendix~\ref{app:pipeline_steps}).
\textsc{inter-annotator agreement} gates deployment: four independent LLM
annotators label stratified held-out traces under a five-phase
deliberation protocol, over up to five rounds of five traces each; the
taxonomy is accepted once mean pairwise area-level agreement reaches
$\kappa \ge 0.75$ with a coverage floor of $0.70$, and a failed round
triggers merge/add/relabel edits and re-runs
(Appendix~\ref{app:iaa_gate}).}
    \label{fig:pipeline}
\end{figure}

The taxonomy generator converts a pool of execution traces into a validated
code set through four phases: \textsc{analysis}, \textsc{curation},
\textsc{consolidation}, and \textsc{inter-annotator agreement}
(Figure~\ref{fig:pipeline}). The split is deliberate. \textsc{analysis} first
extracts the domain context, discovered agent roles, and recurring behavioral
signals from the traces, so that code generation is grounded in observed
behavior rather than generic failure templates. \textsc{curation} then
proposes candidate codes separately per axis. \textsc{consolidation} removes
cross-axis duplicates under strict boundary rules, repairs naming violations,
rejects placeholder or unsupported codes, and enforces per-role and
per-subdomain coverage. In our implementation, these three phases are realized through the
eight-step induction pipeline itemized in
Appendix~\ref{app:pipeline_steps}, which also provides the prompts and
per-step checklists. The numbered steps are not in one-to-one
correspondence with LLM calls: several steps use multiple complementary
calls, Axis~B considers all discovered active roles jointly, and quality
checks can trigger conditional repairs. The subsequent agreement gate
adds its own annotation and revision calls. Taxonomy construction is
therefore a one-time, variable-call cost per target system, amortized
across every downstream procedure and trace that later consumes it
(cost accounting in Appendix~\ref{app:cost_accounting}).

\textsc{inter-annotator agreement} is the deployment gate (full protocol in
Appendix~\ref{app:iaa_gate}). Four LLM annotators independently label the same
held-out trajectories under five sequential phases: independent error
discovery, error reconciliation under a 2-of-4 quorum, failure typing into the
A/B/C axes, code assignment with format validation, and code-level
deliberation bounded at two rounds, while a shared knowledge base of decision
rules, anchor examples, and tracked confusion pairs accumulates across rounds.
Each round labels five stratified held-out traces; up to five rounds run,
terminating early once agreement, computed as mean pairwise Cohen's $\kappa$
over the four annotators, reaches $0.75$ at the failure-area level with a
coverage floor of $0.70$. When the target is not
met, the gate proposes targeted taxonomy edits (merging overlapping codes,
adding codes for uncovered failures, relabeling codes whose boundaries have
drifted) and re-runs on the revised taxonomy.
This gate certifies that the taxonomy is \emph{consistently applicable}; it
does not certify that its codes are correct. Correctness is assessed
externally, against expert human annotations, in \Cref{sec:trail_main}.

\textbf{Online refinement:} The gate's edit operators also serve taxonomy maintenance after deployment.
Because a taxonomy describes the system whose traces produced it, any consumer
whose target system changes can invoke an \emph{online refinement} round: the
current taxonomy is replayed against a pool of recent traces and merge, add,
and relabel edits are proposed, exactly as in the gate's failure loop.
Refinement is thus a property of the taxonomy lifecycle rather than of any one
integration; \Cref{sec:mutation} instantiates its trigger for search, where
the target system changes fastest. Each refinement event adds a bounded number of calls on top of this one-time cost (Appendix~\ref{app:cost_accounting}).

\subsection{Applying the taxonomy}
\label{sec:applying_taxonomy}

Once gated, the taxonomy is exercised by taxonomy-conditioned consumers. The
conditioned reader inspects whatever trace evidence the surrounding procedure
exposes (a completed trajectory, a batch of failed runs, or a mid-run
checkpoint summary) and returns fired codes with supporting evidence. Both
the evidence granularity and the role of the output are properties of the
consumer, not of the taxonomy: the same code set supplies aggregate diagnoses
for a mutation prompt in search (\Cref{sec:mutation}), next-step guidance at
runtime checkpoints (\Cref{sec:runtime_method}), and per-code selection
criteria over completed trajectories (\Cref{sec:verification_method}).
Consumers are also not read-only users of the artifact: whenever the target
system or its trace distribution shifts, any of them can invoke the online
refinement round of \Cref{sec:taxonomy} to keep the vocabulary current. The
next three subsections specify each consumer.

\subsection{\methodname{} for agent-system search}
\label{sec:mutation}

The first integration setting is \emph{agent-system search}: improving the
next generation of agents by rewriting prompts, multi-agent topologies, or
surrounding tooling. We build on AdaEvolve~\citep{adaevolve2025}. Given a
seed agentic system and a task evaluator, AdaEvolve iteratively mutates the
system's source
code to improve evaluator scores, maintaining a population of candidate
variants and selecting which to mutate next based on prior performance.
\methodname{} plugs into this loop at the mutation step. Instead of mutating
from score-only feedback, the mutator receives a taxonomy-based diagnosis of
the parent's execution traces: recurring failure codes, quoted evidence, and
cross-problem patterns that explain why the parent failed (e.g., on
OlympiadBench, \texttt{B.6~Coordinator\_Aggregation\_Mismatch} fires with
quoted evidence that aggregated results conflict due to divergent
\texttt{FINAL ANSWER} lines across branches; representative full diagnoses in
Appendix~\ref{app:olympiad_mechanism}). Here the online refinement round of
\Cref{sec:taxonomy} triggers on stagnation of the evaluator score and operates
over the post-warmup trace pool, so the vocabulary co-evolves with the system
it describes: search changes the system, the system's new traces revise the
taxonomy, and the revised taxonomy shapes the next round of search. Warmup
length, stagnation window, and minimum refinement interval are in
Appendix~\ref{app:implementation} (Table~\ref{tab:hyperparams}); results are
in \Cref{sec:evolution_results}.

\subsection{\methodname{} for runtime monitoring}
\label{sec:runtime_method}

The second integration setting is \emph{runtime monitoring}: surfacing failure
modes during an ongoing trajectory rather than after it ends. The induced taxonomy enters the acting agent's context, and the agent itself is the taxonomy-conditioned reader, reviewing its own recent trace at declared checkpoints. \methodname{}-based reflection is anchored to the system's known failure modes, at no additional LLM calls beyond the agent's own. Delivery follows the harness's capabilities: where the harness exposes mid-run hooks, the taxonomy attaches as a drop-in runtime skill (\textbf{Claude Code} fires checkpoints after tool calls, sub-agent completion, or task completion); where it does not, checkpoints are declared in the agent's instructions and enforced at submission (\textbf{SWE-agent}). Checkpoints come in three kinds: observable checkpoints tied to environment events, introspective checkpoints tied to agent state transitions, and a mandatory pre-submission check; the number and placement differ based on the harness. A check that surfaces fired codes can trigger a bounded repair loop of at most three attempts, after which the agent must report any remaining issues rather than claim success. Per-harness checkpoint schedules and repair-gate details appear in Appendix~\ref{app:runtime_details}; results are
in \Cref{sec:runtime_results}.

\subsection{\methodname{} for trajectory selection}
\label{sec:verification_method}

The third integration setting is \emph{trajectory selection}: given $N$
independent agent trajectories on the same task and no ground-truth checker,
choose the trajectory that actually succeeded \citep{snell2024scaling}. We
refer to the artifact that consumes an adaptive taxonomy in this regime as
\methodname{}-Judge: a verifier that converts induced failure codes into
criteria for judging completed trajectories. The contribution is the criteria
supplied by the induced taxonomy, not the surrounding LLM-as-a-Verifier
pipeline~\citep{kwok2026llmverifier}.

\methodname{}-Judge turns each induced code into a verification criterion
matched to how that failure shows up in a trace: semantic failure modes become
LLM scoring prompts (``does this trajectory exhibit
\texttt{Defensive\_Pivoting}?''), while failures with structural signatures
become inexpensive heuristic checks (trace length for premature termination,
retry counts for repetitive loops). A per-harness forward selector under
leave-one-task-out cross-validation then retains only the most discriminative
criteria. The pool of candidate criteria follows from the induced taxonomy
rather than being fixed in advance; a different target system yields a
different pool (cost and implementation details in
Appendix~\ref{app:cost_accounting}).
Because selection routes the vocabulary through this learned machinery,
its gains reflect both the codes and the selector that consumes them; search
and runtime monitoring, where codes are consumed directly, carry the cleanest
vocabulary claims. Experimental results are in \Cref{sec:terminalbench}.

\section{Experiments}
\label{sec:experiments}

We evaluate whether induced failure taxonomies improve the trace-consuming procedures that use them. Each integration setting tests its own thesis about where the taxonomy earns its keep. \emph{Agent-system search} (\Cref{sec:evolution_results}): taxonomy-coded diagnoses in the mutation prompt produce better post-search architectures than free-form reflection. \emph{Runtime monitoring} (\Cref{sec:runtime_results}): a failure vocabulary consumed at runtime checkpoints raises resolution both on a harness that natively self-verifies and on one where reflection must be prompted. \emph{Trajectory selection} (\Cref{sec:terminalbench}): fired codes discriminate passing from failing trajectories even when routed through a learned selection substrate. Together these support one claim: an induced failure vocabulary is reusable structure for improvement procedures, not a single mechanism. The artifact-level properties of the vocabularies themselves (compactness, human faithfulness, adaptivity) are established separately in \Cref{sec:taxonomy_analysis}.

\subsection{Setup}
\label{sec:experimental_setup}

\paragraph{Benchmarks and integration settings.}
We evaluate three integration settings. For \emph{agent-system search}, we use Frontier-CS~\citep{mang2025frontiercs}, OlympiadBench, MMLU-Pro~\citep{wang2024mmlupro}, TheoremQA~\citep{chen2023theoremqa}, and DROP~\citep{dua2019drop}. For \emph{runtime monitoring}, we use SWE-bench Verified Mini under two
harnesses, Claude Code and SWE-agent~\citep{yang2024sweagent}. We additionally
report a single-run cross-domain transfer study on OfficeQA Pro in
Appendix~\ref{app:officeqa_runtime}. For \emph{trajectory selection}, we use Terminal-Bench~2.0~\citep{merrill2026terminal}, where each task has five pre-recorded trajectories.

\paragraph{Models.}
Evolution experiments use AdaEvolve~\citep{adaevolve2025} as the shared search backbone, with GPT-5.4-mini used for taxonomy generation, mutation, solver agents, and the free-form reflection baseline. Runtime integration uses Claude Haiku 4.5 in the Claude Code harness and GPT-5 in the SWE-agent harness. Terminal-Bench selection uses GPT-5.4 as the verifier for all methods.

\paragraph{Matched comparisons.}
Our comparisons hold the surrounding procedure fixed and vary only the source of failure feedback. Search uses the same AdaEvolve backbone, seed architecture, and iteration number; within each runtime harness, variants use the same solver, instance set, and evaluation protocol; selection uses the same trajectory pools and verifier pipeline. This isolates whether replacing free-form or fixed failure feedback with induced taxonomy feedback changes the downstream procedure. Full implementation details, prompts, seeds, and ablations are in Appendix~\ref{app:implementation}.

\subsection{Agent-system search across reasoning benchmarks}
\label{sec:evolution_results}

The thesis of this experiment: coded diagnoses give the mutation operator better targets than free text. Both arms share the AdaEvolve backbone, seed architecture, and compute budget; the only difference is the failure feedback injected into the mutation prompt (\Cref{sec:mutation}). The baseline receives AdaEvolve's default free-form LLM reflection; \methodname{} receives a taxonomy-coded diagnosis of the parent's failed traces: fired codes, quoted evidence, and cross-problem patterns. 

\begin{table}[t]
\caption{\textbf{Post-search accuracy per benchmark.} \methodname{} adds taxonomy-coded failure feedback to the mutation prompt; LLM Guidance is the default AdaEvolve mutation loop. \emph{Pre-Evol} is the seed architecture's score before evolution. TheoremQA and DROP report means over the multi-seed replications of Appendix~\ref{app:search_multiseed} ($n{=}4$ and $n{=}5$ seeds); MMLU-Pro is a two-seed mean; Frontier-CS and OlympiadBench are single runs. Per-benchmark experimental details are in Appendix~\ref{app:implementation}.}
\label{tab:evolution_results}
\centering
\small
\begin{tabular}{lcccc}
\toprule
Benchmark & Problems & Pre-Evolution & LLM Guidance & \methodname{} \\
\midrule
Frontier-CS (132 unseen) & 132 & $20.8\%$ & 26.0\% & \textbf{32.7\%} \\
OlympiadBench (655 held-out) & 655 & 84.6\% & 87.9\% & \textbf{91.9\%} \\
MMLU-Pro (STEM QA) & 40 & 21.3\% & 35.0\% & \textbf{42.5\%} \\
TheoremQA (math) & 30 & 39.0\% & 60.0\% & \textbf{65.0\%} \\
DROP (discrete reasoning) & 30 & 80.3\% & 88.2\% & \textbf{91.7\%} \\
\bottomrule
\end{tabular}
\end{table}

\begin{figure}[h]
\centering
\includegraphics[width=0.82\linewidth]{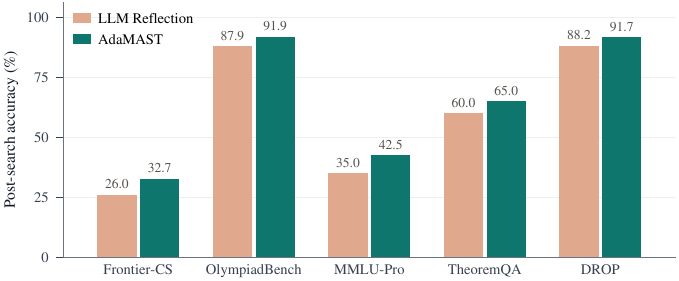}
\caption{\textbf{Search gains across benchmarks.} Post-search accuracy of \methodname{} and the LLM Reflection baseline in \Cref{tab:evolution_results}, across competitive programming, mathematical reasoning, STEM QA, and discrete reasoning. \methodname{} improves on all five benchmarks, by $+3.5$ to $+7.5$\,pp.}
\label{fig:search_gain_summary}
\end{figure}

\paragraph{Search improves across domains.}
Taxonomy-guided mutation outperforms LLM Reflection on all five benchmarks (\Cref{fig:search_gain_summary}): $+6.7$\,pp on Frontier-CS, $+4.0$\,pp on OlympiadBench, $+7.5$\,pp on MMLU-Pro, $+5.0$\,pp on TheoremQA, and $+3.5$\,pp on DROP. The largest benchmark, OlympiadBench, provides the cleanest held-out estimate: the \methodname{}-evolved architecture reaches $91.9\%$ on 655 held-out problems, compared with $87.9\%$ for free-form reflection and $89.5\%$ for a third arm, run only on this benchmark, whose search is guided by the fixed MAST checklist. The small-benchmark seed details and auxiliary search ablations are reported in Appendix~\ref{app:search_ablations}.

\begin{figure*}[t]
\centering
\includegraphics[width=\linewidth]{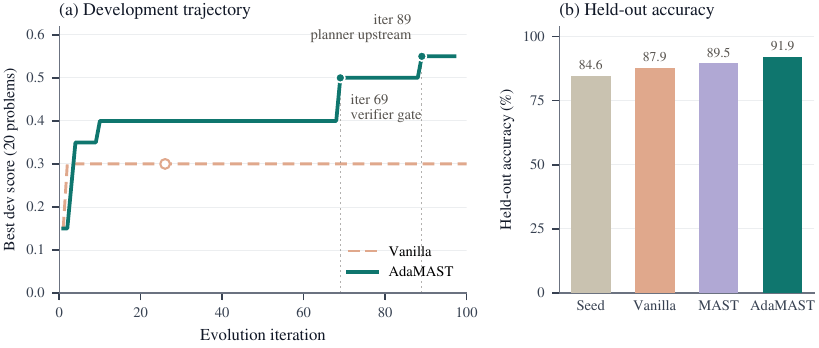}
\caption{\textbf{OlympiadBench search mechanism.} \textbf{(a)} Best dev score per iteration for the vanilla baseline and \methodname{}; both share the AdaEvolve substrate and seed architecture, and only the mutation feedback differs. The open circle marks the vanilla run's exhaustion at iteration 26 (dev score $0.30$). \textbf{(b)} Held-out accuracy on 655 problems across the seed, vanilla, MAST-guided, and \methodname{}-guided architectures. We read this figure as a within-run mechanism trace, not an independent causal ablation.}
\label{fig:olympiad_curve}
\end{figure*}

\paragraph{Case study: evolution on OlympiadBench.}
OlympiadBench has the largest held-out set and the longest evolution trace, so we use it to look inside the search. \Cref{fig:olympiad_curve}(a) compares the \methodname{} run with the \emph{vanilla} run (same backbone and seed, no taxonomy feedback). Vanilla saturates at iteration 26 with dev score $0.30$; \methodname{} reaches $0.40$ early, then jumps to $0.50$ at iteration 69 and $0.55$ at iteration 89, and the gains carry to held-out accuracy (\Cref{fig:olympiad_curve}(b)).

Both jumps are code-driven (\Cref{tab:breakthroughs}): the codes fired on the parent map onto the architecture edits in the child, most directly at the second breakthrough, where B.6 (coordinator aggregation mismatch) prompts the promotion of verification from a binary gate to a score boost. And the codes are signal, not decoration: they fire disproportionately on breakthrough parents, and all existed in the taxonomy before the jumps, so the gains come from the mutator acting on established codes. Judge outputs, base rates, and architecture diffs are in Appendix~\ref{app:olympiad_mechanism}.

\begin{takeawaybox}
\noindent\textbf{Takeaway.} \textit{Taxonomy-coded diagnoses in the mutation prompt beat free-form reflection on all five benchmarks, spanning competitive programming, mathematical reasoning, STEM QA, and discrete reasoning, with gains of $+3.5$ to $+7.5$\,pp under the same search backbone and budget.}
\end{takeawaybox}

\begin{table}[h!]
\centering
\small
\caption{\textbf{Code-driven architectural changes on OlympiadBench.} At each breakthrough, the codes fired on the parent iteration identify recurring failure patterns, which the mutation operator turns into concrete architecture edits.}
\label{tab:breakthroughs}
\begin{tabular}{@{}lp{4.1cm}p{4.6cm}c@{}}
\toprule
Breakthrough & Fired codes (parent) & Architecture mutation & Dev score \\
\midrule
iter 67 $\rightarrow$ 69 & A.3 instruction compliance; B.3 refiner stagnation; C.2 edge-case omission & final-answer verifier; tournament solver pool; edge-case checks in solver prompt & $0.40 \rightarrow 0.50$ \\
\addlinespace
iter 79 $\rightarrow$ 89 & A.6 verifier mismatch; B.6 coordinator aggregation; C.3 over-general prerequisites & planner moved upstream; verifier promoted to score boost; problem-class assumption prompts & $0.50 \rightarrow 0.55$ \\
\bottomrule
\end{tabular}
\end{table}

\subsection{Runtime monitoring on SWE-bench Verified Mini}
\label{sec:runtime_results}

Our primary runtime evaluation tests \methodname{}-based reflection on
SWE-bench Verified Mini under two harnesses, \textbf{Claude Code}~\citep{anthropic_claude_code} (Claude Haiku 4.5 solver) and \textbf{SWE-agent}~\citep{yang2024sweagent} (GPT-5 solver). The checkpoint interaction is the same in both: the agent pauses, inspects its recent steps for the failure modes named by whatever anchors the checkpoint, and repairs any it finds before proceeding (\Cref{sec:runtime_method}); in the \methodname{} arms, the anchor is the induced taxonomy.

The harnesses differ in two native properties. First, checkpoint delivery: Claude Code exposes hooks that fire automatically mid-run (after a tool call, when a sub-agent finishes, at task completion), so the taxonomy attaches as a drop-in runtime skill; SWE-agent has no such enforcement logic, so checkpoints are declared in the agent's instructions and enforced by a submit-gate (Appendix~\ref{app:runtime_details}). Second, native self-verification: Claude Code verifies its own edits unprompted as part of its loop, while SWE-agent triggers reflection only when prompted to. The two harnesses are therefore separate case studies, not a controlled pair—different solvers, different scaffolds, no number compared across them; all controlled comparisons are within a harness. Each study carries its own thesis: on Claude Code, whether anchoring an agent that already self-verifies to its induced failure vocabulary further resolves it; on SWE-agent, where reflection must be added wholesale, what content the added reflection should carry. The SWE-agent arms form a ladder for that question: Base (no reflection), Reflexion (prompted free-text reflection at the checkpoints), MAST~\citep{cemri2025mast} (the same checkpoints anchored to the fixed 14-code checklist), and \methodname{} (the same checkpoints anchored to the induced vocabulary); the three reflection arms share the identical in-prompt scaffold and differ only in content. On Claude Code, comparisons are against vanilla Base and the published 14-code MAST checklist delivered through the identical runtime-skill integration. Checkpoint schedules are in Appendix~\ref{app:runtime_details}.

\begin{table}[t]
\centering
\caption{\textbf{Runtime integration on SWE-bench Verified Mini.} Resolution rate per feedback arm under the two harnesses. Claude Code (Haiku 4.5 solver): average of three seeds of 50 instances each (150 sessions per arm). SWE-agent (GPT-5 solver): one seed of 50 instances. Reflexion runs only under SWE-agent, where reflection must be prompted. The two harnesses are separate case studies; comparisons are within a column. Protocols and per-code firing analysis are in Appendix~\ref{app:runtime_details}.}
\label{tab:swebench_cc}\label{tab:swebench_sweagent}
\begin{tabular}{lcc}
\toprule
Variant & Claude Code & SWE-agent \\
\midrule
Base & 64.0\% & 50\% \\
Reflexion ~\citep{shinn2023reflexion} & -- & 60\% \\
MAST~\citep{cemri2025mast} & 67.3\% & 68\% \\
\methodname{} & \textbf{70.7\%} & \textbf{70\%} \\
\bottomrule
\end{tabular}
\end{table}

\paragraph{Claude Code: an anchored vocabulary raises resolution on a harness that can already verify its work.}
\methodname{} resolves more instances than Base on every seed (\Cref{tab:swebench_cc}), with paired per-seed gaps of $+10/+8/+2$\%, and edges MAST, delivered through the identical skill integration, by $+3.3$\% in the aggregate. Post-hoc code-firing analysis is consistent with the gain being specific to the induced vocabulary rather than a generic checklist effect: in this harness, the induced 32-code taxonomy's role axis names single-agent \emph{phases} (Edit/Plan/Verify) rather than multi-agent roles, and the induced vocabulary primarily reduces verification-phase failures, including \textbf{B.8} \emph{Verify ignored import or syntax errors} and \textbf{B.6} \emph{Plan skips verification loop}, while MAST's reductions concentrate on broader patch-footprint errors (\textbf{B.3} \emph{Edit overbroad patch footprint}). At the instance level the gain is concentrated where it counts: \methodname{} converts instances that resolve only occasionally under Base into reliable resolves (Appendix~\ref{app:runtime_details}).

\paragraph{SWE-agent: under an identical checkpoint scaffold, the induced vocabulary is the strongest anchor.} On SWE-agent (\Cref{tab:swebench_sweagent}), \methodname{}-based reflection resolves $35/50$ instances ($70\%$): $+20$\% over Base, $+10$\% over free-text Reflexion, and $+2$\% over the fixed MAST checklist. Because the three reflection arms share the identical scaffold and differ only in what anchors the check, the ordering isolates content: adding prompted reflection at all recovers $+10$\% over Base, anchoring it to a categorical vocabulary adds a further $+8$\% (MAST over Reflexion), and the induced, harness-specific vocabulary adds another $+2$\%.

\paragraph{Cross-domain transfer to OfficeQA.}
To test whether the runtime integration extends beyond software repair, we
also evaluate it on 133 hard, oracle-parsed OfficeQA Pro tasks under the
Claude Code harness with a Haiku 4.5 solver. The frozen 15-code taxonomy,
induced from 50 baseline transcripts, resolves 69 of 133 tasks, compared with
59 of 133 for Base. Because this is a single-run transfer study without
budget-matched accounting, we treat it as evidence that the integration
operates in a new domain rather than as a confirmatory effectiveness
comparison. The complete protocol is reported in
Appendix~\ref{app:officeqa_runtime}.

\begin{takeawaybox}
\noindent\textbf{Takeaway.} \textit{Consumed at runtime checkpoints, as a drop-in skill on Claude Code, or in-prompt on SWE-agent, the induced vocabulary raises resolution on both harnesses: $64.0\%\!\to\!70.7\%$ on Claude Code and $50\%\!\to\!70\%$ on SWE-agent, above free-text reflection ($60\%$) and the fixed MAST checklist ($68\%$).}
\end{takeawaybox}

\subsection{Trajectory selection on Terminal-Bench 2.0}
\label{sec:terminalbench}

Finally, we evaluate the trajectory-selection setting of \Cref{sec:verification_method} on Terminal-Bench~2.0, choosing among each task's five pre-recorded trajectories with no ground-truth checker at selection time. The thesis here is structural: fired failure codes carry enough signal to rank completed trajectories. We use three harnesses: terminus-2~\citep{merrill2026terminal}, claude-code~\citep{anthropic_claude_code}, and ForgeCode~\citep{tailcallhq2026forgecode}. The trajectory pool is fixed across methods, so any difference comes from the selector rather than new agent samples.

\begin{table}[t]
\caption{\textbf{Best-of-$5$ selection accuracy on Terminal-Bench~2.0.} The MAST row substitutes the published 14-code MAST taxonomy~\citep{cemri2025mast} into the same verifier pipeline used by \methodname{}-Judge; only the failure vocabulary changes.}
\label{tab:terminalbench}
\centering
\small
\begin{tabular}{lccc}
\toprule
Method & terminus-2 & claude-code & ForgeCode \\
\midrule
Pass@1 (no selection) & 61.8\% & 57.5\% & 81.8\% \\
LLM-as-a-Verifier~\citep{kwok2026llmverifier} & 71.2\% & 61.2\% & 86.5\% \\
MAST~\citep{cemri2025mast} & 68.5\% & 69.0\% & 88.8\% \\
\methodname{}-Judge & \textbf{73.0\%} & \textbf{72.4\%} & \textbf{89.9\%} \\
Best-of-5 oracle & 77.5\% & 80.5\% & 89.9\% \\
\bottomrule
\end{tabular}
\end{table}

\methodname{}-Judge improves over Pass@1 by $+11.2$, $+14.9$, and $+8.1$\,pp across the three harnesses, and improves over MAST by $+4.5$, $+3.4$, and $+1.1$\,pp. 
On ForgeCode, both MAST and \methodname{} approach the oracle ceiling, so the comparison is less informative. As a sanity check against the taxonomy overfitting the evaluation pool, a held-out 5-fold validation regenerates the taxonomy on train folds and evaluates on held-out task categories; held-out accuracy meets or exceeds the same-pool result on the two non-saturated configurations. We report this as a robustness check rather than a confirmatory result.

\begin{figure}[t]
\centering
\includegraphics[width=0.95\linewidth]{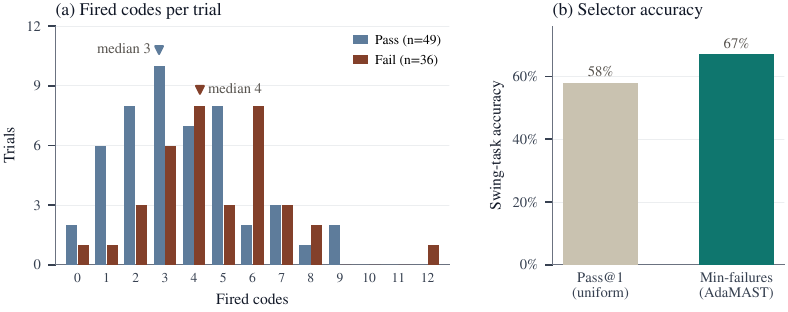}
\caption{\textbf{Fired codes and trajectory outcome.} The ForgeCode \methodname{} taxonomy is applied to every trial of the 17 ForgeCode swing tasks. \textbf{(a)} Distribution of fired codes for passing versus failing trajectories; triangles mark the medians (3 for passing, 4 for failing). \textbf{(b)} Expected swing-task accuracy of a fewest-fired-codes selector versus uniform selection ($58\% \rightarrow 67\%$, $+9$\,pp).}
\label{fig:tb_trial_failures}
\end{figure}

\paragraph{Failure codes discriminate passing from failing trajectories.}
Taxonomy-guided selection presupposes that fired codes track true failure. To test this directly, we apply the ForgeCode \methodname{} taxonomy to all 85 trials of the 17 ForgeCode \emph{swing tasks} (tasks whose five trials mix pass and fail, the only ones where selection can change the outcome) and count fired codes (\Cref{fig:tb_trial_failures}). Passing trajectories fire fewer codes than failing trajectories: median 3 versus 4. A selector that picks the trial with the fewest fired codes reaches 67\% expected swing-task accuracy, compared with 58\% for uniform Pass@1-style selection. The failure-mode signal is informative before adding the learned verifier substrate.

\paragraph{Selection is substrate-sensitive.}
Selection routes taxonomy criteria through the feature selector and heuristic verifier substrate of \Cref{sec:verification_method}, so the gain reflects both the induced vocabulary and the selection pipeline that consumes it. We therefore treat Terminal-Bench as evidence that failure codes are useful for selection, and reserve the cleanest claims about adaptive vocabulary content for the regimes where the taxonomy is consumed directly. The detailed selection ablations, including wrong-domain taxonomies, cheaper verifiers, top-$K$ codes, token-matched context, and cross-verifier replication, are in Appendix~\ref{app:selection_ablations}.

\begin{takeawaybox}
\noindent\textbf{Takeaway.} \textit{Fired codes rank completed trajectories: \methodname{}-Judge improves best-of-$5$ accuracy by $+8.1$ to $+14.9$\,pp over Pass@1, retaining $+3.4$/$+4.5$\,pp over the fixed MAST vocabulary on the two non-saturated harnesses.}
\end{takeawaybox}

\section{Taxonomy Validation and Analysis}
\label{sec:taxonomy_analysis}
The experiments above test whether induced failure vocabularies improve downstream procedures. We now analyze the taxonomies themselves. We ask whether the induced codes align with human failure judgments, whether different domains induce different vocabularies, whether taxonomy-guided search changes the observed failure surface over time, and how much of a trace's failure-relevant content the codes retain relative to their size.

\subsection{Human-anchored validation on TRAIL}
\label{sec:trail_main}

An induced taxonomy is useful only if its codes name failures that are recognizable beyond the LLM that generated them. We test this on TRAIL~\citep{deshpande2025trail}, which contains 117 GAIA-derived traces annotated by four software-engineering experts under a hand-crafted 20-category taxonomy.

\paragraph{Agreement with expert labels.}
A four-LLM panel applies the \methodname{}-induced TRAIL taxonomy using span-grounded prompts, one round of peer deliberation, and majority vote. The panel reaches Cohen's $\kappa{=}0.725$ against the four-expert human gold at the failure-area level (2-of-4 consensus threshold; Table~\ref{tab:trail_bootstrap}). This indicates that the induced codes recover broad human-recognizable failure areas rather than merely forming an internally consistent LLM labeling scheme. Substituting a same-model panel (four GPT-5.4 annotators) for the cross-family panel lowers agreement from $\kappa{=}0.682$ to $0.625$ at the 1-of-4 reporting threshold under an otherwise identical protocol (Appendix~\ref{app:trail}), indicating that at this threshold the alignment is carried by the induced codes rather than by a labeling regime shared within one model family.

\paragraph{Vocabulary isolation.}
To isolate the value of the induced vocabulary itself, we hold the panel and prompts fixed and vary only the taxonomy consumed by the annotators (\Cref{tab:trail_vocab_comparison}). TRAIL's hand-crafted 20-category vocabulary yields area-$\kappa{=}0.516$, while \methodname{}'s induced vocabulary yields area-$\kappa{=}0.682$. Thus, under the same annotation procedure, the induced vocabulary better matches expert-labeled failure areas. Fine-grained leaf-level agreement, prompt details, annotator agreement, and the mapping between induced codes and TRAIL categories are reported in Appendix~\ref{app:trail}.

\begin{table}[t]
\centering
\small
\caption{\textbf{TRAIL faithfulness.} Area-level Cohen's $\kappa$ against the four-expert TRAIL gold. The first two rows hold the annotation panel and prompts fixed and vary only the vocabulary; the third row runs the full \methodname{} protocol on the induced vocabulary.}
\label{tab:trail_vocab_comparison}
\begin{tabular}{llc}
\toprule
Vocabulary & Annotation protocol & Area-$\kappa$ \\
\midrule
TRAIL hand-crafted (20 categories) & matched panel + prompts & 0.516 \\
\methodname{}-induced & matched panel + prompts & 0.682 \\
\methodname{}-induced & span grounding + deliberation & \textbf{0.725} \\
\bottomrule
\end{tabular}
\end{table}

\subsection{Cross-domain taxonomy variation}
\label{sec:taxonomy_variation}

If \methodname{} were merely rediscovering a fixed generic checklist, the code sets induced for different domains would substantially overlap. They do not: across six evolution domains, full code sets overlap at a mean pairwise Jaccard of only $0.14$, so most codes are specific to the agent roles, tools, and task distribution that produced the traces. Projected onto a universal failure backbone, overlap rises to $0.50$: the domains share broad cross-cutting failure families while disagreeing on nearly every concrete code. This is the intended design: the top-level axes stay fixed, and the vocabulary inside them adapts to the target system (pairwise overlap matrices in \Cref{fig:jaccard}, Appendix~\ref{app:cross_domain}).

\subsection{Failure modes over search}
\label{sec:failure_burden}

Does taxonomy-guided search change \emph{what} fails, or only how often? On the OlympiadBench run (the longest recorded evolution trace), it changes what fails: by iteration 92, 12 of the 28 mid-run codes no longer fire and the severity-weighted failure burden drops by 23\%, even though firings per iteration stay roughly constant. Failure modes are retired, and the remaining firings concentrate on harder residual cases. The mix also shifts with the architecture: as evolution adds roles, the role-specific (B) share of firings grows from 39.8\% to 46.9\%, while the domain-specific (C) share falls from 32.0\% to 25.5\% as the solver acquires domain guard rails, and the system-level (A) share stays flat. Finally, the two taxonomy-refinement events (iterations 29 and 78) each precede one of the run's score jumps (iterations 69 and 89). This is not a causal ablation, but it is consistent with the vocabulary participating in the breakthroughs rather than labeling them after the fact. Per-code lifecycles, base rates, and post-hoc failure profiles are in Appendix~\ref{app:olympiad_mechanism}.

\subsection{Codes as a compressed failure representation}
\label{sec:compression}

The premise behind treating a taxonomy as a feedback interface is that named codes carry the failure-relevant content of a trace at a fraction of its size. On 223 traces from the SWE-bench runtime study (\Cref{sec:runtime_results}), coding compresses the failure-relevant content by ${\sim}18\times$ without collapsing the distinctions between traces: all 30 codes fire across the corpus, the median trace fires 5 codes, and $89\%$ of traces retain a unique code signature, so two traces almost never map to the same diagnosis (measurement details in Appendix~\ref{app:functional_compression}).

The compressed artifact also \emph{functionally} substitutes for the traces it summarizes. We freeze an induction pool of $N \le 40$ labeled runs and ask a consumer LLM (Qwen3.5-122B-A10B, Qwen3.5-27B) to predict held-out run success given either the verbatim pool or the taxonomy induced from it (protocol, all six context conditions, and full grids in Appendix~\ref{app:functional_compression}). On Terminal-Bench, the taxonomy matches the verbatim pool at every $N$ for both consumers, while adding ${\sim}95\times$ fewer context tokens ($1.2$K vs.\ $114$K). On TheoremQA, the taxonomy alone is the strongest condition at every $N$ for the 27B consumer, beating the verbatim pool at $N \ge 10$ ($0.925$ vs.\ $0.850$ at $N{=}20$, $p \le .012$) with ${\sim}2$K tokens against the pool's $19$--$40$K; for the 122B consumer, the taxonomy plus three labeled runs takes that role ($0.917$ at $N{=}20$) while accuracy under the verbatim pool \emph{degrades} as the pool grows ($0.767$ at $N{=}40$, below its $0.850$ no-context baseline). Consumption is most effective when the taxonomy is simply present in context: a paired ablation that forces the consumer to audit the current run code-by-code before deciding \emph{lowers} accuracy ($-0.13$ to $-0.24$, $p \le .006$) by biasing it toward predicting failure even when the run succeeds (Appendix~\ref{app:functional_compression}). A downstream procedure can therefore consume the taxonomy in place of the trace pool at one to two orders of magnitude less context.


\section{Conclusion}
\label{sec:conclusion}
\methodname{} treats an agent system's failure vocabulary as infrastructure: induced once from the system's own traces, validated once, and then consumed by every procedure that would otherwise re-derive its diagnosis in disposable free text. The three consumers we evaluated are deliberately dissimilar (an evolutionary search loop, a runtime self-check, a best-of-$N$ verifier), yet each improved once its feedback was anchored to its system's induced codes, and the artifact earned trust before any downstream gain: compact enough to stand in for traces, faithful to expert judgment, and specific to the system that produced it. The general lesson is that between raw traces and scalar outcomes sits a missing representation: named, evidence-grounded, system-specific failure modes. Agents already generate the evidence of their own failures on every run; the cheapest improvement to an agent system may be to first learn, in its own terms, how it fails. 

\bibliographystyle{abbrvnat}
\bibliography{references}

@inproceedings{cemri2025mast,
  title     = {Why Do Multi-Agent LLM Systems Fail?},
  author    = {Cemri, Mert and Pan, Melissa Z. and Yang, Shuyi and Agrawal, Lakshya A. and Chopra, Bhavya and Tiwari, Rishabh and Keutzer, Kurt and Parameswaran, Aditya and Klein, Dan and Ramchandran, Kannan and Zaharia, Matei and Gonzalez, Joseph E. and Stoica, Ion},
  booktitle = {Advances in Neural Information Processing Systems},
  year      = {2025}
}

@inproceedings{chen2023theoremqa,
  title={Theoremqa: A theorem-driven question answering dataset},
  author={Chen, Wenhu and Yin, Ming and Ku, Max and Lu, Pan and Wan, Yixin and Ma, Xueguang and Xu, Jianyu and Wang, Xinyi and Xia, Tony},
  booktitle={Proceedings of the 2023 Conference on Empirical Methods in Natural Language Processing},
  pages={7889--7901},
  year={2023}
}

@article{adaevolve2025,
  title={Adaevolve: Adaptive llm driven zeroth-order optimization},
  author={Cemri, Mert and Agrawal, Shubham and Gupta, Akshat and Liu, Shu and Cheng, Audrey and Mang, Qiuyang and Naren, Ashwin and Erdogan, Lutfi Eren and Sen, Koushik and Zaharia, Matei and others},
  journal={arXiv preprint arXiv:2602.20133},
  year={2026}
}

@misc{anthropic_claude_code,
  title        = {Claude Code},
  author       = {{Anthropic}},
  year         = {2025},
  howpublished = {\url{https://docs.anthropic.com/en/docs/claude-code/overview}},
  note         = {Accessed: 2026-05-07}
}

@article{merrill2026terminal,
  title={Terminal-bench: Benchmarking agents on hard, realistic tasks in command line interfaces},
  author={Merrill, Mike A and Shaw, Alexander G and Carlini, Nicholas and Li, Boxuan and Raj, Harsh and Bercovich, Ivan and Shi, Lin and Shin, Jeong Yeon and Walshe, Thomas and Buchanan, E Kelly and others},
  journal={arXiv preprint arXiv:2601.11868},
  year={2026}
}

@misc{kwok2026llmverifier,
  title={LLM-as-a-Verifier: A General-Purpose Verification Framework},
  author={Jacky Kwok and Shulu Li and Pranav Atreya and Yuejiang Liu and Yixing Jiang and Chelsea Finn and Marco Pavone and Ion Stoica and Azalia Mirhoseini},
  year={2026},
  eprint={2607.05391},
  archivePrefix={arXiv},
  primaryClass={cs.AI}
}

@article{liu2026evox,
  title={Evox: Meta-evolution for automated discovery},
  author={Liu, Shu and Agarwal, Shubham and Maheswaran, Monishwaran and Cemri, Mert and Li, Zhifei and Mang, Qiuyang and Naren, Ashwin and Boneh, Ethan and Cheng, Audrey and Pan, Melissa Z and others},
  journal={arXiv preprint arXiv:2602.23413},
  year={2026}
}

@article{snell2024scaling,
  title={Scaling llm test-time compute optimally can be more effective than scaling model parameters},
  author={Snell, Charlie and Lee, Jaehoon and Xu, Kelvin and Kumar, Aviral},
  journal={arXiv preprint arXiv:2408.03314},
  year={2024}
}

@article{agrawal2025gepa,
  title={Gepa: Reflective prompt evolution can outperform reinforcement learning},
  author={Agrawal, Lakshya A and Tan, Shangyin and Soylu, Dilara and Ziems, Noah and Khare, Rishi and Opsahl-Ong, Krista and Singhvi, Arnav and Shandilya, Herumb and Ryan, Michael J and Jiang, Meng and others},
  journal={arXiv preprint arXiv:2507.19457},
  year={2025}
}

@article{romeraparedes2024funsearch,
  author    = {Romera-Paredes, Bernardino and Barekatain, Mohammadamin and Novikov, Alexander and others},
  title     = {Mathematical Discoveries from Program Search with Large Language Models},
  journal   = {Nature},
  volume    = {625},
  pages     = {468--475},
  year      = {2024}
}

@article{novikov2025alphaevolve,
  author    = {Novikov, Alexander and Balog, Matej and Chaudhuri, Swarat and others},
  title     = {{AlphaEvolve}: A Coding Agent for Scientific and Algorithmic Discovery},
  journal   = {arXiv preprint arXiv:2506.13131},
  year      = {2025}
}

@inproceedings{hu2025adas,
  author    = {Hu, Shengran and Lu, Cong and Clune, Jeff},
  title     = {Automated Design of Agentic Systems},
  booktitle = {International Conference on Learning Representations (ICLR)},
  year      = {2025}
}

@article{yuan2024evoagent,
  author    = {Yuan, Siyu and others},
  title     = {{EvoAgent}: Towards Automatic Multi-Agent Generation via Evolutionary Algorithms},
  journal   = {arXiv preprint arXiv:2406.14228},
  year      = {2024}
}

@article{yang2024sweagent,
  title={SWE-agent: Agent-Computer Interfaces Enable Automated Software Engineering},
  author={Yang, John and Jimenez, Carlos E. and Wettig, Alexander and Lieret, Kilian and Yao, Shunyu and Narasimhan, Karthik and Press, Ofir},
  journal={arXiv preprint arXiv:2405.15793},
  year={2024}
}

@inproceedings{zhuge2024gptswarm,
  author    = {Zhuge, Mingchen and others},
  title     = {{GPTSwarm}: Language Agents as Optimizable Graphs},
  booktitle = {Proceedings of the International Conference on Machine Learning (ICML)},
  year      = {2024}
}

@inproceedings{lightman2024prm,
  author    = {Lightman, Hunter and Kosaraju, Vineet and Burda, Yura and others},
  title     = {Let's Verify Step by Step},
  booktitle = {International Conference on Learning Representations (ICLR)},
  year      = {2024}
}

@inproceedings{zheng2023llmjudge,
  author    = {Zheng, Lianmin and Chiang, Wei-Lin and Sheng, Ying and others},
  title     = {Judging {LLM}-as-a-Judge with {MT-Bench} and Chatbot Arena},
  booktitle = {Advances in Neural Information Processing Systems (NeurIPS)},
  year      = {2023}
}

@inproceedings{shinn2023reflexion,
  author    = {Shinn, Noah and Cassano, Federico and Gopinath, Ashwin and others},
  title     = {Reflexion: Language Agents with Verbal Reinforcement Learning},
  booktitle = {Advances in Neural Information Processing Systems (NeurIPS)},
  year      = {2023}
}

@inproceedings{khattab2024dspy,
  author    = {Khattab, Omar and others},
  title     = {{DSPy}: Compiling Declarative Language Model Calls into State-of-the-Art Pipelines},
  booktitle = {International Conference on Learning Representations (ICLR)},
  year      = {2024}
}

@inproceedings{madaan2023selfrefine,
  author    = {Madaan, Aman and others},
  title     = {Self-Refine: Iterative Refinement with Self-Feedback},
  booktitle = {Advances in Neural Information Processing Systems (NeurIPS)},
  year      = {2023}
}

@article{pan2026measuringagentsproduction,
  author    = {Pan, Michael Z. and Arabzadeh, Negar and Cogo, Rodrigo and others},
  title     = {Measuring Agents in Production},
  journal   = {arXiv preprint arXiv:2512.04123},
  year      = {2026}
}

@inproceedings{cheng2024trace,
  author    = {Cheng, Ching-An and Nie, Allen and Swaminathan, Adith},
  title     = {Trace is the Next {AutoDiff}: Generative Optimization with Rich Feedback, Execution Traces, and {LLM}s},
  booktitle = {Advances in Neural Information Processing Systems (NeurIPS)},
  year      = {2024}
}

@article{yuksekgonul2024textgrad,
  author    = {Yuksekgonul, Mert and Bianchi, Federico and Boen, Joseph and others},
  title     = {{TextGrad}: Automatic ``Differentiation'' via Text},
  journal   = {arXiv preprint arXiv:2406.07496},
  year      = {2024}
}

@inproceedings{aflow2025,
  author    = {Zhang, Jiayi and Xiang, Jinyu and Yu, Zhaoyang and others},
  title     = {{AFlow}: Automating Agentic Workflow Generation},
  booktitle = {International Conference on Learning Representations (ICLR)},
  year      = {2025}
}

@article{fernando2023promptbreeder,
  author    = {Fernando, Chrisantha and Banarse, Dylan and Michalewski, Henryk and Osindero, Simon and Rockt\"aschel, Tim},
  title     = {{Promptbreeder}: Self-Referential Self-Improvement via Prompt Evolution},
  journal   = {arXiv preprint arXiv:2309.16797},
  year      = {2023}
}

@inproceedings{guo2024evoprompt,
  author    = {Guo, Qingyan and Wang, Rui and Guo, Junliang and others},
  title     = {Connecting Large Language Models with Evolutionary Algorithms Yields Powerful Prompt Optimizers},
  booktitle = {International Conference on Learning Representations (ICLR)},
  year      = {2024}
}

@article{jia2025gpa,
  author  = {Jia, Allison Sihan and Huang, Daniel and Vytla, Nikhil and Yoo, Seung Won Wilson and Choudhury, Nirvika and Sen, Shayak and Mitchell, John C. and Datta, Anupam},
  title   = {What Is Your Agent's {GPA}? A Framework for Evaluating Agent Goal-Plan-Action Alignment},
  journal = {arXiv preprint arXiv:2510.08847},
  year    = {2025}
}

@article{deshpande2025trail,
  author  = {Deshpande, Darshan and Gangal, Varun and Mehta, Hersh and Krishnan, Jitin and Kannappan, Anand and Qian, Rebecca},
  title   = {{TRAIL}: Trace Reasoning and Agentic Issue Localization},
  journal = {arXiv preprint arXiv:2505.08638},
  year    = {2025}
}

@inproceedings{wang2024mmlupro,
  author    = {Wang, Yubo and Ma, Xueguang and Zhang, Ge and Ni, Yuansheng and Chandra, Abhranil and Guo, Shiguang and Ren, Weiming and Arulraj, Aaran and He, Xuan and Jiang, Ziyan and Li, Tianle and Ku, Max and Wang, Kai and Zhuang, Alex and Fan, Rongqi and Yue, Xiang and Chen, Wenhu},
  title     = {{MMLU-Pro}: A More Robust and Challenging Multi-Task Language Understanding Benchmark},
  booktitle = {Advances in Neural Information Processing Systems (NeurIPS), Datasets and Benchmarks Track},
  year      = {2024}
}

@inproceedings{dua2019drop,
  author    = {Dua, Dheeru and Wang, Yizhong and Dasigi, Pradeep and Stanovsky, Gabriel and Singh, Sameer and Gardner, Matt},
  title     = {{DROP}: A Reading Comprehension Benchmark Requiring Discrete Reasoning Over Paragraphs},
  booktitle = {Proceedings of NAACL-HLT},
  pages     = {2368--2378},
  year      = {2019}
}

@misc{tailcallhq2026forgecode,
  author       = {{TailcallHQ}},
  title        = {{ForgeCode}: An Open-Source Multi-Agent Coding Harness},
  year         = {2026},
  howpublished = {\url{https://github.com/tailcallhq/forgecode}; \url{https://forgecode.dev}},
  note         = {Top-ranked open-source agent on Terminal-Bench 2.0 (Pass@1 81.8\% with the Forge harness)}
}

@misc{mang2025frontiercs,
  title         = {{Frontier-CS}: Evolving Challenges for Evolving Intelligence},
  author        = {Mang, Qiuyang and Chai, Wenhao and Li, Zhifei and Mao, Huanzhi and Zhou, Shang and Du, Alexander and Li, Hanchen and Liu, Shu and Chen, Edwin and Wang, Yichuan and others},
  year          = {2025},
  eprint        = {2512.15699},
  archivePrefix = {arXiv},
  primaryClass  = {cs.LG},
  url           = {https://arxiv.org/abs/2512.15699}
}

@inproceedings{eyuboglu2022domino,
  title     = {Domino: Discovering systematic errors with cross-modal embeddings},
  author    = {Eyuboglu, Sabri and Varma, Maya and Saab, Khaled and Delbrouck, Jean-Benoit and Lee-Messer, Christopher and Dunnmon, Jared and Zou, James and R{\'e}, Christopher},
  booktitle = {International Conference on Learning Representations (ICLR)},
  year      = {2022}
}

@inproceedings{ribeiro2020checklist,
  title     = {Beyond accuracy: Behavioral testing of NLP models with {CheckList}},
  author    = {Ribeiro, Marco Tulio and Wu, Tongshuang and Guestrin, Carlos and Singh, Sameer},
  booktitle = {Association for Computational Linguistics (ACL)},
  year      = {2020}
}

@article{mcaleese2024criticgpt,
  title   = {LLM critics help catch {LLM} bugs},
  author  = {McAleese, Nat and Pokorny, Rai Michael and Uribe, Juan Felipe Cer{\'o}n and Nitishinskaya, Evgenia and Trebacz, Maja and Leike, Jan},
  journal = {arXiv preprint arXiv:2407.00215},
  year    = {2024}
}

@article{zhang2024generativeverifier,
  title   = {Generative verifiers: Reward modeling as next-token prediction},
  author  = {Zhang, Lunjun and Hosseini, Arian and Bansal, Hritik and Kazemi, Mehran and Kumar, Aviral and Agarwal, Rishabh},
  journal = {arXiv preprint arXiv:2408.15240},
  year    = {2024}
}

@article{zhang2025whowhen,
  title={Which Agent Causes Task Failures and When? On Automated Failure Attribution of {LLM} Multi-Agent Systems},
  author={Zhang, Shaokun and Yin, Ming and Zhang, Jieyu and Liu, Jiale and Han, Zhiguang and Zhang, Jingyang and Li, Beibin and Wang, Chi and Wang, Huazheng and Chen, Yiran and Wu, Qingyun},
  journal={arXiv preprint arXiv:2505.00212},
  year={2025}
}

\newpage
\appendix

\section{Generated Taxonomy Examples}
\label{app:taxonomy_examples}

This appendix presents representative induced taxonomies. For each benchmark, we report a short context paragraph and the top-five failure codes by firing frequency. The full machine-generated catalogues, with code descriptions and per-iteration provenance, are released alongside the run artifacts.

\subsection{TheoremQA Taxonomy (13 codes)}

The TheoremQA evolution pipeline evaluates 30 graduate-level math problems with a flat solver-verifier architecture. Because the architecture has no differentiated roles, the induced taxonomy contains no role-specific B-codes. The dominant failure, \texttt{Premature\_Reasoning\_Truncation}, names the pattern where the solver halts a multi-step derivation mid-calculation and emits an intermediate quantity as the final answer. The remaining high-frequency failures concern context exhaustion, unit handling, numerical precision, and instruction compliance.

\begin{figure}[H]
\centering
\includegraphics[width=0.92\linewidth]{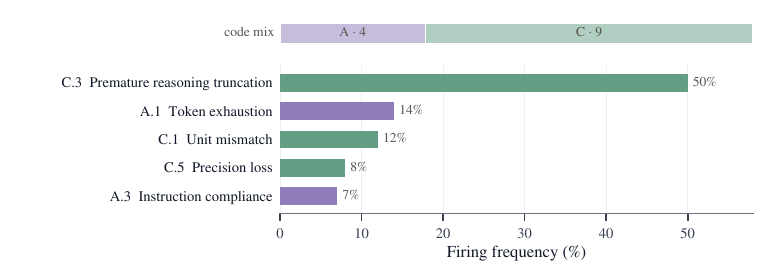}
\caption{Top induced failure codes for TheoremQA (13 codes: 4 system, 9 domain). Bars show firing frequency; colors indicate the \methodname{} axis (\textcolor{adamastpurple}{A} system, \textcolor{adamastgreen}{C} domain). Failures concentrate on truncated derivations and unit/precision drift.}
\label{fig:taxonomy_theoremqa_topcodes}
\end{figure}
\FloatBarrier

\subsection{Frontier-CS Taxonomy (25 codes)}

The Frontier-CS run produces a 25-code taxonomy (5A/13B/7C) for a competitive-programming multi-agent system spanning solver, reviewer, and classifier roles. Algorithm-mismatch, budget exhaustion, and stagnation are the most common failure modes; role-specific failures concentrate on solver bias toward familiar algorithms.

\begin{figure}[H]
\centering
\includegraphics[width=0.92\linewidth]{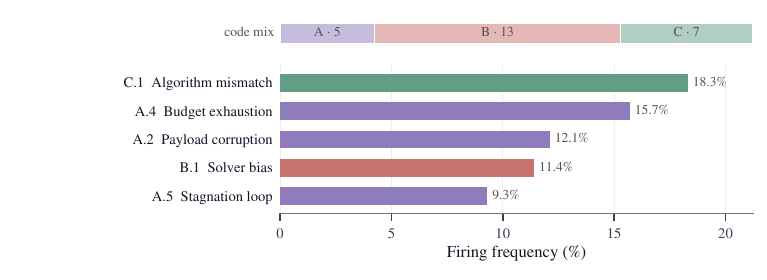}
\caption{Top induced failure codes for Frontier-CS (25 codes: 5 system, 13 role, 7 domain). Bars show firing frequency; colors indicate the \methodname{} axis (\textcolor{adamastpurple}{A} system, \textcolor{adamastcoral}{B} role, \textcolor{adamastgreen}{C} domain). Failures concentrate on algorithm-family mismatch, solver bias, and budget exhaustion.}
\label{fig:taxonomy_frontiercs_topcodes}
\end{figure}
\FloatBarrier

\subsection{OlympiadBench Taxonomy (36 codes)}

The OlympiadBench taxonomy is refined across the 97-iteration run; the initial pre-refinement version contains 36 codes (16A/6B/14C); later snapshots are smaller, as refinement events retire more codes than they add. The most-fired codes target output-format compliance, solver reasoning quality, and combinatorial counting errors. Refinement events at iterations 29, 78, and 92 introduce new codes for failure modes the evolving architecture newly surfaces. (Table~\ref{tab:cross_domain} reports the smaller final post-refinement snapshot of this taxonomy, 16 codes; the 36-code count here refers to the initial pre-refinement version.)

\begin{table}[H]
\centering
\small
\caption{Top five OlympiadBench failure codes by firing frequency in the mid-run taxonomy.}
\label{tab:olympiad_taxonomy_top5}
\begin{tabular}{llrp{5cm}}
\toprule
Code & Name & Freq. & Impact \\
\midrule
A.3 & Instruction\_Compliance\_Violation & 100\% & Final-answer format violations \\
C.2 & Edge\_Case\_Omission & 94.4\% & Edge cases enumerated but not validated \\
C.4 & Algebraic\_Manipulation\_Gap & 91.7\% & Incomplete or flawed algebraic steps \\
A.1 & Stagnation\_Anomaly & 91.7\% & Non-convergent iteration patterns \\
C.5 & Combinatorics\_Counting\_Misreasoning & 88.9\% & Naive counting decompositions \\
\bottomrule
\end{tabular}
\end{table}
\FloatBarrier

\subsection{Terminal-Bench Taxonomy}

The Terminal-Bench taxonomy yields scoring criteria that target failure modes specific to the benchmark's task categories (security, machine learning, system administration, data science). Each induced code becomes an actionable verification criterion for \methodname{}-Judge.

\begin{table}[H]
\centering
\small
\caption{Representative \methodname{}-derived verification criteria on Terminal-Bench~2.0.}
\label{tab:terminalbench_taxonomy}
\begin{tabular}{llp{6cm}}
\toprule
Category & Criterion & What it catches \\
\midrule
System & Premature\_Termination & Trajectory ends before task completion \\
Security & Defensive\_Pivoting & Subtle shift from offensive to defensive posture \\
ML & Training\_Stall\_Detection & No error message, just timeout \\
Sys-Admin & Service\_State\_Verification & systemctl fails in Docker, agent unaware \\
Data Science & Output\_Format\_Alignment & Column-name case sensitivity \\
\bottomrule
\end{tabular}
\end{table}
\FloatBarrier

\subsection{SWE-bench Taxonomy (30 codes)}

The SWE-bench taxonomy is induced from SWE-agent trajectories on Verified Mini and contains 30 codes (8A/11B/11C). The dominant failures concern ad-hoc validation, misread root causes, and unclean patch contents.

\begin{figure}[H]
\centering
\includegraphics[width=0.92\linewidth]{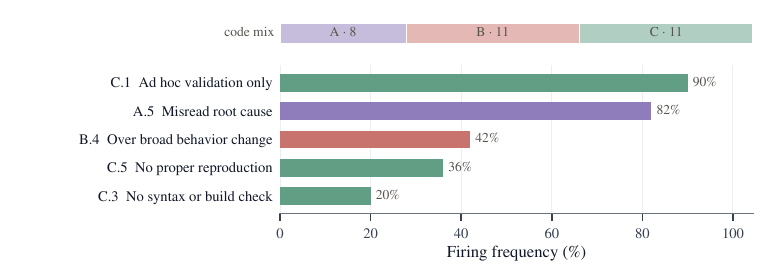}
\caption{Top induced failure codes for SWE-bench under the SWE-agent harness (30 codes: 8 system, 11 role, 11 domain). Bars show firing frequency; colors indicate the \methodname{} axis (\textcolor{adamastpurple}{A} system, \textcolor{adamastcoral}{B} role, \textcolor{adamastgreen}{C} domain). Dominant failures concern shallow validation and misread root causes.}
\label{fig:taxonomy_swebench_topcodes}
\end{figure}
\FloatBarrier

\subsection{Cross-domain taxonomy comparison}
\label{app:cross_domain}

Table~\ref{tab:cross_domain} compares the induced taxonomies across the seven domains for which we induce them (the six Jaccard-comparison domains plus OlympiadBench). Three patterns emerge. C-codes are entirely domain-specific (algorithm mismatch in competitive programming, physical-law violation in STEM QA, negotiation-leverage failure in economic simulation). B-codes appear only in multi-agent systems with specialized roles. A-codes form a partial universal backbone of cross-cutting failures. Two related overlap statistics appear in the paper and measure different things: Table~\ref{tab:jaccard_matrix} restricts to A-codes only and reports mean $J=0.47$ over the six Jaccard-comparison domains, while the universal-backbone projection of \Cref{sec:taxonomy_variation} projects \emph{full} taxonomies onto a fixed set of cross-cutting failure tokens and yields $0.50$ over the same six domains.

\begin{figure}[H]
\centering
\includegraphics[width=0.85\linewidth]{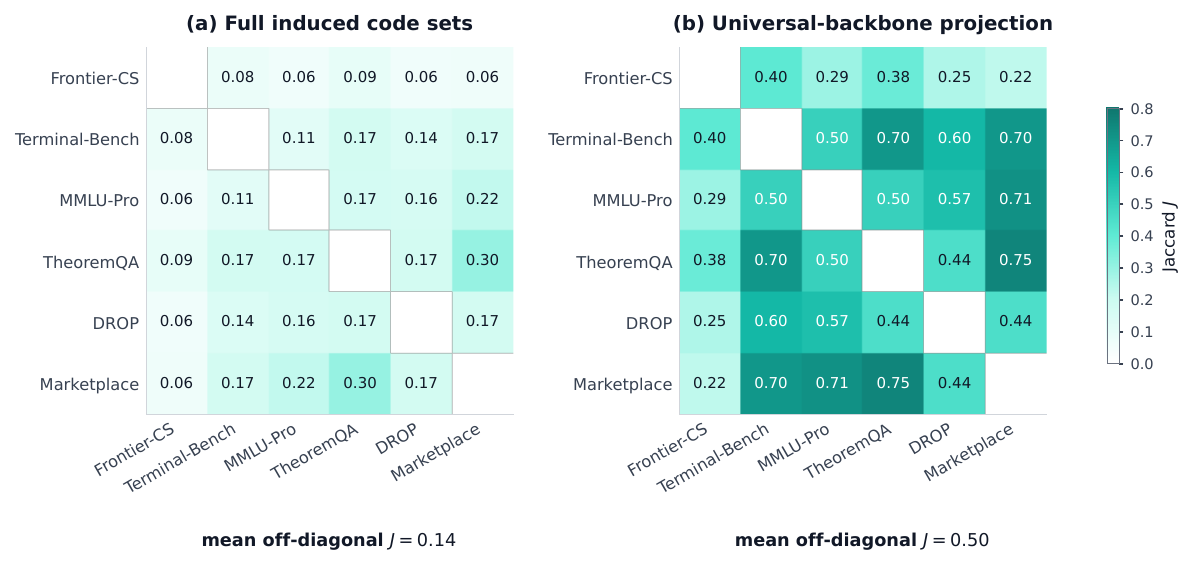}
\caption{\textbf{Cross-domain code overlap.} Pairwise Jaccard between the induced taxonomies for six evolution domains. \textbf{(a)} Full code sets (mean off-diagonal $0.14$). \textbf{(b)} Each domain projected onto a universal failure backbone: a fixed set of cross-cutting failure tokens such as context exhaustion, looping, and format violation (mean $0.50$).}
\label{fig:jaccard}
\end{figure}

\begin{table}[H]
\caption{Generated taxonomy size per domain (final snapshot per run; the OlympiadBench row is the post-refinement taxonomy, cf.\ the 36-code pre-refinement version above). The A column counts system-level codes, B counts role-specific codes, and C counts domain-reasoning codes.}
\label{tab:cross_domain}
\centering
\small
\begin{tabular}{lcccp{4.8cm}}
\toprule
Domain & A & B & C & Representative C-code \\
\midrule
Frontier-CS & 5 & 13 & 7 & Complexity\_Class\_Constraint\_Mismatch \\
OlympiadBench & 6 & 5 & 5 & Combinatorics\_Counting\_Misreasoning \\
Terminal-Bench & 10 & 16 & 18 & Defensive\_Pivoting \\
MMLU-Pro & 6 & 0 & 10 & Physical\_Law\_Violation \\
TheoremQA & 4 & 0 & 9 & Premature\_Reasoning\_Truncation \\
DROP & 4 & 0 & 10 & Arithmetic\_Operation\_Error \\
Marketplace & 6 & 2 & 6 & Negotiation\_Leverage\_Failure \\
\bottomrule
\end{tabular}
\end{table}

\subsection{Long-tail structure of induced codes}
\label{app:long_tail}

\begin{figure}[h]
\centering
\includegraphics[width=\linewidth]{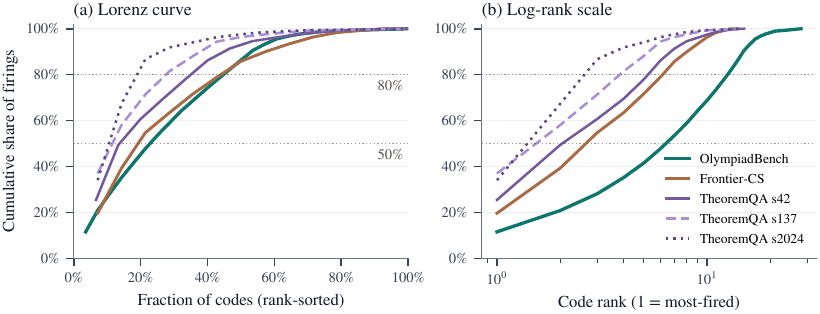}
\caption{\textbf{Long-tail structure of induced taxonomies.} \textbf{(a)} Lorenz curves: cumulative share of code firings against the rank-sorted fraction of codes; the top $\sim$20\% of codes account for the bulk of firings on every run. \textbf{(b)} Same data on a log-rank axis. The shape is consistent across OlympiadBench, Frontier-CS, and three TheoremQA seeds, indicating that the concentration is a property of the induced vocabulary rather than of a single run.}
\label{fig:failure_pareto}
\end{figure}

Induced taxonomies are concentrated: a handful of codes account for most firings, with a long tail of low-frequency codes. \Cref{fig:failure_pareto} plots cumulative share of firings against the rank-sorted codes (Lorenz curve and log-rank) across the OlympiadBench, Frontier-CS, and three TheoremQA seeds. The top $\sim$20\% of codes already cover $\sim$60--80\% of firings depending on the run, and the curves have similar shapes across domains and seeds.

\begin{table}[H]
\centering
\small
\caption{Per-pair Jaccard similarity over A-code structural backbones across six evolution domains. Mean $J = 0.47$. The QA-style domains cluster tightly, while Frontier-CS sits furthest because its A-codes are dominated by tool-and-dependency failures.}
\label{tab:jaccard_matrix}
\begin{tabular}{lcccccc}
\toprule
 & Frontier-CS & Terminal-B. & MMLU-Pro & TheoremQA & DROP & Marketplace \\
\midrule
Frontier-CS    & 1.00 & --   & --   & --   & --   & --   \\
Terminal-Bench & 0.25 & 1.00 & --   & --   & --   & --   \\
MMLU-Pro       & 0.17 & 0.63 & 1.00 & --   & --   & --   \\
TheoremQA      & 0.20 & 0.50 & 0.50 & 1.00 & --   & --   \\
DROP           & 0.00 & 0.63 & 0.67 & 0.50 & 1.00 & --   \\
Marketplace    & 0.17 & 0.63 & 1.00 & 0.50 & 0.67 & 1.00 \\
\bottomrule
\end{tabular}
\end{table}

\section{Taxonomy Generation and Implementation Details}
\label{app:implementation}

\subsection{Taxonomy generation pipeline}
\label{app:pipeline_steps}

Taxonomy construction comprises an eight-step induction pipeline followed by an inter-annotator agreement gate. The induction pipeline covers analysis, curation, consolidation, and internal quality checks. The subsequent agreement stage applies the draft taxonomy to held-out traces with independent LLM annotators and accepts it only if mean pairwise Cohen's $\kappa$ reaches $0.75$ at the failure-area level with a coverage floor of $0.70$.

The eight numbered steps are not in one-to-one correspondence with LLM calls. \textbf{Step~1 (Domain analysis)} characterizes the domain, subdomains, terminology, and common error patterns from a stratified trace sample. \textbf{Step~2 (Role and topology discovery)} extracts agent names through regex preprocessing and uses LLM analysis to infer free-form functional roles from their observed behavior and identify the architecture topology. Common roles such as solver, checker, refiner, and coordinator are provided only as examples; the generator may discover other roles when warranted by the traces. An intervening LLM-free signal extractor identifies behavioral patterns such as truncation, looping, refusal, and tool errors. \textbf{Steps~3--5} generate system-level, role-specific, and domain-specific codes, respectively. Each category uses two complementary generation passes followed by within-category consolidation. Category B considers all discovered active roles jointly, and every resulting code is tagged with an \texttt{applies\_to\_role} field. \textbf{Step~6 (Cross-category deduplication)} removes semantic duplicates across the three axes. \textbf{Step~7 (Structural validation)} checks category placement, naming, and role attribution. \textbf{Step~8 (Quality and coverage check)} detects remaining overlaps, structural issues, and coverage gaps and invokes targeted repairs when needed.

\subsubsection*{Taxonomy-generation audit}

We analyze 203 complete taxonomy-generation records spanning multiple target systems, domains, and deployment settings. Each record contains the full sequence of outputs from Steps~3 through~8 and uses the same taxonomy-generation pipeline version. We measure how code inventories change during generation and, where recorded, during the first refinement round.

\begin{table}[h]
\centering
\small
\caption{\textbf{Inventory changes during taxonomy generation and first-round refinement.} Reduction is computed separately for each taxonomy; negative values denote net code addition.}
\label{tab:taxonomy_generation_audit}
\begin{tabular}{lrrrr}
\toprule
Stage & IQR & Median & Mean & Full range \\
\midrule
\multicolumn{5}{l}{\textit{Complete generation records ($n=203$)}} \\
Raw generated inventory & 21--30 & 24 & 26.5 & 8--63 \\
After deduplication & 20--28 & 22 & 25.0 & 8--57 \\
After structural validation & 19--27 & 22 & 24.0 & 8--57 \\
Final inventory & 17--24 & 20 & 22.2 & 6--54 \\
Per-taxonomy raw-to-final reduction
    & 8.7--22.7\% & 14.9\% & 16.4\% & $-25.0\%$ to $+65.7\%$ \\
\midrule
\multicolumn{5}{l}{\textit{First refinement rounds ($n=32$)}} \\
Final inventory before refinement & 19--29 & 22.5 & 24.3 & 9--37 \\
Inventory after first refinement & 18--27 & 21 & 21.5 & 6--35 \\
Per-taxonomy reduction
    & 0--21.8\% & 4.7\% & 10.7\% & $-47.6\%$ to $+71.4\%$ \\
\bottomrule
\end{tabular}
\end{table}

Across the 203 generation records, the median inventory decreases from 24 raw codes to 20 final codes. The pooled inventory decreases from 5,379 to 4,498 codes, a reduction of 16.4\%. The successive transitions reduce the pooled inventory by 5.6\% from raw generation to deduplication, 4.1\% from deduplication to structural validation, and 7.6\% from structural validation to the final output. At the individual-taxonomy level, 181 inventories shrink, 7 remain unchanged, and 15 grow. The pipeline therefore generally consolidates the generated vocabulary without enforcing a fixed inventory size.

Among the 32 records with a measurable first refinement round, the median inventory decreases from 22.5 to 21 codes. Twenty-one inventories shrink, nine remain unchanged, and two grow. Refinement can therefore merge or retire existing codes while also adding failures newly observed as the target system changes.

Together, these measurements show that generation and refinement can both consolidate and expand the vocabulary rather than forcing a fixed code count.

\subsection{Inter-annotator agreement gate}
\label{app:iaa_gate}

After Step~8, the draft taxonomy is validated against held-out traces through a multi-phase deliberative annotation protocol. Four LLM annotators independently label the same trajectories under five sequential phases: (1)~independent error discovery, (2)~error reconciliation with a 2-of-4 quorum, (3)~high-level failure typing into Type~A/B/C, (4)~code assignment with format validation, and (5)~code-level deliberation bounded at two rounds. A shared knowledge base of decision rules, anchor examples, and confusion-pair tracking accumulates across rounds. The protocol targets a mean pairwise Cohen's $\kappa \geq 0.75$ across the four annotators at the area level, with a coverage floor of $0.70$. Each round samples five stratified traces; up to five rounds run, terminating early when both targets are met. When the $\kappa$ target is not met, the loop proposes targeted edits (merge, add, relabel) and re-runs against the revised taxonomy.

\subsection{Models and prompts}
\label{app:models_prompts}

Taxonomy generation, mutation, and free-form reflection use GPT-5.4-mini. Terminal-Bench selection uses GPT-5.4 as the verifier for all methods, with all three agents based on Claude Opus~4.6. Runtime integration uses GPT-5 as the SWE-agent solver and Claude Haiku~4.5 as the Claude Code solver. TRAIL validation uses a four-model panel of Claude Opus~4.7, GPT-5.4, Gemini-2.5-Pro, and GPT-5.4-mini. Full prompt text for each pipeline step is released alongside the run artifacts.

\paragraph{Hyperparameters.}
\begin{table}[h]
\centering
\small
\caption{Hyperparameter settings used in the search-side experiments.}
\label{tab:hyperparams}
\begin{tabular}{lrl}
\toprule
Parameter & Value & Rationale \\
\midrule
Warmup rounds ($N_{\text{warm}}$) & 5 & Balances trace pool quality vs.\ compute \\
Stagnation window ($w$) & 10 iter. & One full population evaluation cycle \\
Stagnation threshold ($\epsilon$) & 0.10 & Above per-iteration noise floor \\
Min refine interval ($r_{\min}$) & 10 iter. & Prevents back-to-back refinement \\
Population size & 20 & Standard for island-model GA \\
Number of islands & 3 & Diversity without fragmentation \\
Migration interval & 8 iter. & Allow island divergence before sharing \\
Feedback codes injected ($K$) & 5--7 & Top codes by frequency \\
Random seeds & 42, 123, 456 & Standard reproducible seeds \\
\bottomrule
\end{tabular}
\end{table}

\paragraph{Seed architecture.}
Every evolution run begins from the same minimal three-agent seed (analyzer, solver, verifier) with a sequential topology and a solve-verify-refine loop. Starting from the same seed across LLM-reflection and \methodname{} runs ensures the comparison isolates the contribution of taxonomy-coded feedback, not an architectural prior.

\subsection{Benchmark splits and evaluation budgets}
\label{app:benchmark_splits}

\paragraph{Search benchmarks.}
Frontier-CS uses 132 unseen competitive-programming problems. OlympiadBench uses a 20-problem dev set during search and a 655-problem held-out test set for final evaluation. MMLU-Pro uses a 40-problem stratified STEM subset. TheoremQA uses 30 graduate-level math problems. DROP uses 30 discrete-reasoning problems. Each run uses an AdaEvolve budget of 40--100 iterations.

\paragraph{Runtime benchmark.}
SWE-bench Verified Mini contains 50 instances. Each SWE-agent run uses a per-instance limit of 75 LLM calls, a \$4 per-task cost ceiling, and a \$200 per-run cost ceiling at temperature 0 with greedy decoding.

\paragraph{Selection benchmark.}
Terminal-Bench~2.0 contains 89 tasks. We use 17--35 swing tasks per configuration (tasks with mixed pass/fail across 5 trials) under leave-one-task-out cross-validation.

\paragraph{TRAIL.}
TRAIL contains 117 GAIA-derived agent traces hand-annotated by four software-engineering experts under a 20-category taxonomy. No TRAIL labels are used to induce the \methodname{} taxonomy.

\subsection{Cost accounting}
\label{app:cost_accounting}

\methodname{} adds LLM-call overhead with three recurring cost shapes (per-task for selection, per-iteration for search, and per-checkpoint for runtime monitoring), plus a one-time taxonomy-construction cost for each target system and domain. Let $C_{\mathrm{gen}}=C_{\mathrm{draft}}+C_{\mathrm{IAA}}$, where $C_{\mathrm{draft}}$ covers the eight-step induction pipeline and $C_{\mathrm{IAA}}$ covers agreement validation and any resulting revision rounds. For a standard run with nonempty A, B, and C inventories, the reference implementation makes 17 draft-stage calls before conditional repairs or optional code capping: one for domain analysis, two for structure and role discovery, three for each category, one for cross-category deduplication, one for structural validation, and three initial overlap checks. The total is variable because quality repairs and agreement rounds run only when required. This one-time cost is amortized whenever the resulting taxonomy is reused.

\paragraph{Search-time overhead.} Let $N$ be the number of evolution iterations, $T$ the tasks per iteration, $K$ the LLM calls per task during evaluation. Evaluation cost is $C_{\text{eval}} = N \cdot T \cdot K$ calls. Two overhead components: (i) a two-pass judge that fires only on novel parents, $C_{\text{judge}} = \rho \cdot N \cdot (T+1)$ with $\rho \approx 0.4$ in our runs; (ii) refinement events, $C_{\text{refine}}=\sum_{j=1}^{r} c_j$, where $c_j$ is the variable call cost of the edits and validation performed during refinement event $j$. The dominant ratio of overhead to evaluation cost in the large-$N$ limit is $\rho / K$, which is small for any non-trivial multi-agent architecture. No overhead at deployment: the loop ships a Python program with no \methodname{} machinery in the runtime loop.

\paragraph{Runtime overhead.} The \methodname{} runtime integration carries the taxonomy in the agent's context and adds zero extra LLM calls.

\paragraph{Selection overhead.} The 20 heuristic features are regex-based with negligible LLM cost. LLM-scoring overhead per task is at most $6N + \binom{N}{2}$ calls in the worst case where $N$ is candidate trajectories; typically far less because the per-agent forward selector retains 5--10 of the 27 candidate features, often heuristic-heavy.

\paragraph{Per-arm runtime cost.} The four-arm SWE-agent runtime study costs \$24--28 per arm at 25 parallel workers; the three-arm Claude Code study costs \$25--27 per arm (\Cref{tab:swe_cost_cc}). The \methodname{} arm additionally incurs the one-time taxonomy-induction and agreement-gating cost described above. Because this cost is paid when producing the frozen taxonomy and amortized across its subsequent uses, we report it separately from task-level runtime costs.

\section{Search Ablations and Replications}
\label{app:search_ablations}

This appendix reports the search-side controls omitted from the main text. The main result in \Cref{sec:evolution_results} holds the AdaEvolve substrate, seed architecture, and compute budget fixed while changing only the feedback source injected into the mutation prompt. These additional experiments test multi-seed stability, taxonomy transfer across domains, sample efficiency of induction, and replication under a different model backbone.

\subsection{Multi-seed stability on small benchmarks}
\label{app:search_multiseed}

The main-table TheoremQA and DROP results use small evaluation sets, so we repeated the comparison across additional seeds. Each paired comparison uses the same benchmark, AdaEvolve substrate, seed architecture, and iteration budget, changing only whether the mutation prompt receives taxonomy-coded feedback.

\begin{table}[h]
\centering
\small
\caption{\textbf{Multi-seed stability on small search benchmarks.} Each entry reports post-search performance after AdaEvolve. These small benchmarks are used as directional replications of the main search result.}
\label{tab:search_multiseed}
\begin{tabular}{lccc}
\toprule
Benchmark & LLM Reflection & \methodname{} & Per-seed gaps \\
\midrule
TheoremQA ($n{=}4$ seeds)
& $60.0 \pm 6.1$
& $\mathbf{65.0 \pm 4.3}$
& $+3.3,+3.3,+10.0,+3.3$ \\
DROP ($n{=}5$ seeds)
& $88.2 \pm 3.0$
& $\mathbf{91.7 \pm 2.1}$
& $0.0,+7.3,+2.5,+5.2,+2.5$ \\
\bottomrule
\end{tabular}
\end{table}

TheoremQA improves in all four seeds; DROP improves in four of five seeds and ties in one. The small-benchmark replication is consistent with the main search result.

\subsection{Wrong-domain taxonomy transfer in search}
\label{app:search_wrong_domain}

We test whether replacing the matched taxonomy with a wrong-domain taxonomy changes search performance. Search is the regime where the mutation prompt consumes the taxonomy text directly, so vocabulary mismatch can affect proposed architecture edits.

\begin{table}[h]
\centering
\small
\caption{\textbf{Wrong-domain taxonomy transfer in search.} MMLU-Pro post-search accuracy under matched and wrong-domain taxonomy feedback.}
\label{tab:search_wrong_domain}
\begin{tabular}{lcc}
\toprule
Feedback taxonomy & Mean accuracy & Mean gap vs matched \\
\midrule
Matched MMLU-Pro taxonomy & $\mathbf{37.8\%}$ & -- \\
Frontier-CS taxonomy & $33.3\%$ & $-4.5$\,pp \\
TheoremQA taxonomy & $35.6\%$ & $-2.2$\,pp \\
\bottomrule
\end{tabular}
\end{table}

The matched taxonomy is best on average. The direction supports the regime distinction in the discussion: vocabulary content matters more when the downstream procedure consumes the taxonomy directly.

\subsection{Sample efficiency of taxonomy induction}
\label{app:sample_efficiency}

We test how many warmup traces are needed before an induced taxonomy becomes useful as search feedback. We induce TheoremQA taxonomies from $N \in \{5,10,20,40\}$ traces sampled with a deterministic seed, freeze each taxonomy, and run 40 iterations of AdaEvolve with refinement disabled.

\begin{table}[h]
\centering
\small
\caption{\textbf{Sample efficiency of taxonomy induction.} TheoremQA post-search accuracy as a function of the number of traces used to induce the taxonomy. Even small trace pools produce useful feedback.}
\label{tab:sample_efficiency}
\begin{tabular}{lcc}
\toprule
Induction traces & Taxonomy size & Post-search accuracy \\
\midrule
$N{=}5$  & 13 codes & $60.0\%$ \\
$N{=}10$ & 17 codes & $60.0\%$ \\
$N{=}20$ & 16 codes & $50.0\%$ \\
$N{=}40$ & 21 codes & $56.7\%$ \\
Vanilla baseline & -- & $\sim 46.7\%$ \\
\bottomrule
\end{tabular}
\end{table}

The curve is not monotonic at this sample size, so we do not claim a scaling law in the number of induction traces. The useful conclusion is simpler: a small trace pool is already sufficient to produce feedback that improves over the vanilla search baseline.

\subsection{Cross-model search replication}
\label{app:cross_model_search}

We replicate the search comparison under a different model backbone (Gemini-3-flash-preview) with three seeds per cell and 40 AdaEvolve iterations. This is a robustness check; several small benchmarks saturate near the ceiling.

\begin{table}[h]
\centering
\small
\caption{\textbf{Cross-model search replication.} Paired \methodname{}-vs-vanilla gaps under a Gemini-3-flash-preview solver/judge/mutator backbone. The direction of the \methodname{} effect is non-negative across all reported benchmarks.}
\label{tab:cross_model_search}
\begin{tabular}{lccc}
\toprule
Benchmark & Vanilla & \methodname{} & Mean gap \\
\midrule
TheoremQA & $78.9\%$ & $\mathbf{81.1\%}$ & $+2.2$\,pp \\
MMLU-Pro & $90.0\%$ & $\mathbf{100.0\%}$ & $+10.0$\,pp \\
DROP & $98.8\%$ & $\mathbf{99.4\%}$ & $+0.6$\,pp \\
OlympiadBench & $71.7\%$ & $\mathbf{86.7\%}$ & $+15.0$\,pp \\
\bottomrule
\end{tabular}
\end{table}

The magnitudes differ from the main table because the solver model and evaluation regime differ. We include this replication to show that the direction of the \methodname{} effect is not tied to a single model family.

\section{Selection Ablations}
\label{app:selection_ablations}

This appendix reports the Terminal-Bench selection ablations. The main text reports the headline selection result and the fired-code discrimination analysis. These ablations characterize the selection substrate. Terminal-Bench selection is mediated by a feature selector and heuristic features, which makes selection a deployment-pattern study rather than an isolated test of vocabulary wording.

\subsection{Wrong-domain taxonomy transfer}
\label{app:wrong_domain_selection}

We replace the matched Terminal-Bench taxonomy with taxonomies induced from other domains and re-run the same \methodname{}-Judge pipeline. All auto-generated rows use the same 6-prompt prosecutor+health substrate, 20 heuristic features, forward feature selector, and leave-one-task-out cross-validation; only the taxonomy text changes.

\begin{table}[h]
\centering
\small
\caption{\textbf{Wrong-domain taxonomy transfer on Terminal-Bench~2.0.} Full-benchmark accuracy when the matched taxonomy is replaced by a taxonomy generated for another benchmark. This ablation shows that the Terminal-Bench selector can recover useful signal even when taxonomy text is mismatched, because the feature selector can downweight unhelpful taxonomy prompts and rely on general trace features.}
\label{tab:wrong_domain_verif}
\begin{tabular}{lccc}
\toprule
Source taxonomy & terminus-2 & claude-code & ForgeCode$^*$ \\
\midrule
Matched Terminal-Bench taxonomy & $\mathbf{73.0\%}$ & $\mathbf{72.4\%}$ & $87.6\%$ \\
Frontier-CS taxonomy & $73.0\%$ & $70.1\%$ & $87.6\%$ \\
OlympiadBench taxonomy & $74.2\%$ & $72.4\%$ & $\mathbf{88.8\%}$ \\
MMLU-Pro taxonomy & $71.9\%$ & $72.4\%$ & $87.6\%$ \\
\midrule
MAST-14 & $68.5\%$ & $69.0\%$ & $\mathbf{88.8\%}$ \\
\bottomrule
\end{tabular}
\par\vspace{2pt}
\footnotesize $^*$ ForgeCode uses the 26-feature apples-to-apples variant that omits the legacy MAST pairwise tiebreaker.
\end{table}

\subsection{Cheaper verifier}
\label{app:cheaper_verifier}

We test whether \methodname{}-Judge depends on the strong GPT-5.4 verifier by replacing it with GPT-5.4-mini while holding the matched taxonomy, 6-prompt substrate, 20 heuristic features, forward selector, and LOTO cross-validation fixed.

\begin{table}[h]
\centering
\small
\caption{\textbf{Cheaper-verifier ablation.} Replacing GPT-5.4 with GPT-5.4-mini leaves selection accuracy close to the original result on the two non-saturated configurations.}
\label{tab:cheaper_verifier}
\begin{tabular}{lcc}
\toprule
Verifier & terminus-2 & claude-code \\
\midrule
GPT-5.4 & $73.0\%$ & $\mathbf{72.4\%}$ \\
GPT-5.4-mini & $\mathbf{74.2\%}$ & $71.3\%$ \\
\bottomrule
\end{tabular}
\end{table}

This suggests that the selection pipeline does not require the largest verifier model once the failure-oriented features are available.

\subsection{Top-$K$ taxonomy truncation}
\label{app:topk_selection}

We truncate the matched 44-code Terminal-Bench taxonomy to its first $K{=}5$ codes, preserving the A/B/C proportions, and re-run \methodname{}-Judge on terminus-2.

\begin{table}[h]
\centering
\small
\caption{\textbf{Top-$K$ code truncation.} On terminus-2, a truncated 5-code taxonomy matches the full-taxonomy selector.}
\label{tab:topk_selection}
\begin{tabular}{lcc}
\toprule
Taxonomy size & Accuracy & Selected features \\
\midrule
Full taxonomy & $73.0\%$ & mixed heuristic/LLM features \\
$K{=}5$ codes & $73.0\%$ & \texttt{n\_success}, \texttt{llm\_local}, \texttt{n\_error}, \texttt{n\_killed} \\
\bottomrule
\end{tabular}
\end{table}

This shows that the Terminal-Bench selector is not primarily sensitive to the full code-list length. It can recover signal from a small subset of codes plus generic trace features.

\subsection{Token-matched context}
\label{app:token_matched_selection}

We test whether structured taxonomy text is necessary for the selection-side prompt context by comparing three context types: \methodname{} taxonomy text, a one-shot prose summary of failure signals, and raw trace excerpts. We evaluate each under token budgets of 100, 500, and 2000 tokens on terminus-2.

\begin{table}[h]
\centering
\small
\caption{\textbf{Token-matched context ablation for selection.} In this selector, taxonomy text, prose summaries, and raw excerpts fall within a narrow range, indicating that the surrounding selection substrate contributes substantially to the result.}
\label{tab:token_matched_selection}
\begin{tabular}{lccc}
\toprule
Context type & 100 tokens & 500 tokens & 2000 tokens \\
\midrule
\methodname{} taxonomy & $72.3\%$ & $72.3\%$ & $72.3\%$ \\
Prose summary & $72.3\%$ & $72.3\%$ & $72.3\%$ \\
Raw trace excerpts & $75.3\%$ & $72.3\%$ & $75.0\%$ \\
\bottomrule
\end{tabular}
\end{table}

We treat this as a deployment-pattern analysis rather than an isolated test of vocabulary wording.

\subsection{Cross-verifier replication}
\label{app:cross_verifier_selection}

We repeat the wrong-domain substitution using Claude Sonnet 4.6 as the verifier with single-letter A--T scoring rather than GPT-5.4 logprobs. The mean swing-accuracy change across the wrong-source cells is approximately $+0.1$\,pp. Pairwise Pearson correlations between per-trial LLM scores across the five taxonomies are high ($r=0.89$--$0.93$ on the claude-code swing pool), and the forward selector picks at most two LLM features per cell. The cross-verifier result is consistent with the wrong-domain transfer above: in this selection substrate, the learned selector and heuristic features carry most of the marginal effect of taxonomy wording.

\section{Compression Measurement and Functional Substitution}
\label{app:functional_compression}

\subsection{Compression ratios and entropy}

We quantify compression on 223 traces from the SWE-bench runtime study by comparing each trace against its judge-assigned code set. Amortizing the taxonomy text (the codebook, shipped once) across the corpus, the coded representation compresses the failure surface by ${\sim}18\times$ under gzip (${\sim}38\times$ in raw characters, ${\sim}25\times$ under a token-cost proxy); treating each trace standalone, the median per-trace reduction is $9\times$.

The reduction is also visible in the empirical entropy of the realized text. Judge rationales have substantially higher surface diversity than the coded representation: unigram entropy drops from \(8.36\) bits/token for judge free text to \(4.29\) bits/token for bare codes and \(5.58\) bits/token for the conservative standalone encoding. Length-matched bootstrap samples from judge rationales remain near \(7.4\)--\(8.1\) bits/token, indicating that the entropy reduction is not merely an artifact of the coded traces being shorter. Per-trace compression is flat across feedback arms and across passing and failing traces, indicating a property of the representation rather than of any one experimental condition.

\subsection{Functional substitution: protocol and full results}

We test \emph{functional} substitution directly (\Cref{tab:functional_compression,tab:functional_tb}): freeze an induction pool $P$ of $N \in \{5,10,20,40\}$ labeled runs, give a consumer LLM one of six contexts --- no failure context, $P$ verbatim, three random traces from $P$, a length-matched free-text summary of $P$, the taxonomy induced from $P$, or the taxonomy plus three labeled runs from $P$ (two passing, one failing) --- and ask it to predict whether a held-out run of the same system succeeds (task- or problem-disjoint from $P$; probability read from the decision-token logprobs; taxonomies induced with a Llama-3.3-70B backbone; consumers Qwen3.5-122B-A10B and Qwen3.5-27B; two pool seeds; Terminal-Bench pools drawn from ForgeCode runs and TheoremQA pools from the search traces of \Cref{sec:evolution_results}). On Terminal-Bench no taxonomy-vs-pool contrast reaches significance at any $N$ for either consumer ($0.82$ vs.\ $0.84$ at $N{=}40$ for the 27B), while the taxonomy adds ${\sim}95\times$ fewer context tokens over the no-context baseline ($1.2$K vs.\ $114$K). On TheoremQA the taxonomy alone is the strongest condition at every $N$ for the 27B consumer, beating the verbatim pool at $N \ge 10$ ($0.917$--$0.925$ vs.\ $0.825$--$0.858$, $p \le .012$) with ${\sim}2$K tokens against the pool's $19$--$40$K; for the 122B consumer the taxonomy-plus-labeled-runs hybrid takes that role ($0.917$ at $N{=}20$, $0.900$ at $N{=}40$; beats the pool at $p{=}.039$ and $p<10^{-4}$), and three random labeled traces are similarly competitive ($0.908$ at $N{=}40$) --- outcome labels carry calibration. Verbatim experience does not scale: the 122B consumer's accuracy under the raw pool \emph{degrades} as the pool grows ($0.867 \to 0.767$ from $N{=}5$ to $N{=}40$), ending below its own no-context baseline ($0.850$).

\emph{How the taxonomy is consumed matters, in the opposite direction one might expect.} In a paired protocol ablation, we forced the consumer to apply the taxonomy code-by-code to the current run --- citing fired codes with trace snippets --- before emitting its verdict. This \emph{lowers} TheoremQA balanced accuracy for the taxonomy conditions by $0.13$--$0.24$ ($p \le .006$ at every $N$, both consumers) while leaving the verbatim-pool condition unharmed or improved ($+0.125$ at $N{=}40$ for the 122B, $p{=}.0007$): the enforced audit turns the consumer into a process critic that verdicts failure on real intermediate pathologies even when the run succeeds, collapsing pass recall while ranking is preserved (AUC $0.87$--$0.92$). Passive in-context provision, not enforced checklist application, is the effective consumption mode for current-generation consumers. With earlier-generation consumers (Qwen3-235B, Qwen2.5-72B) the qualitative picture was the same on Terminal-Bench and for the weaker consumer on TheoremQA, but the strongest consumer then extracted additional calibration from the verbatim pool at $N{=}40$ --- an edge that disappears with the current consumers.

\begin{table}[t]
\caption{\textbf{Functional substitution on TheoremQA, all pool sizes}
(balanced accuracy, pooled over two pool seeds; decision-token protocol).
Tokens = mean prompt tokens at $N{=}40$. Bold = best per column. The induced
taxonomy is the strongest condition at every $N$ for the 27B consumer
(vs.\ verbatim pool: $p \le .012$ at $N \ge 10$); for the 122B, the
taxonomy${+}$labeled-runs hybrid and random labeled traces lead, and the
verbatim pool degrades as it grows.}
\label{tab:functional_compression}
\centering
\small
\begin{tabular}{lccccccccc}
\toprule
 & & \multicolumn{4}{c}{Qwen3.5-27B} & \multicolumn{4}{c}{Qwen3.5-122B-A10B} \\
\cmidrule(lr){3-6}\cmidrule(lr){7-10}
Context & tokens & $N{=}5$ & $N{=}10$ & $N{=}20$ & $N{=}40$ & $N{=}5$ & $N{=}10$ & $N{=}20$ & $N{=}40$ \\
\midrule
no failure context        & 1.2K  & \multicolumn{4}{c}{0.900} & \multicolumn{4}{c}{0.850} \\
3 random traces           & 3.9K  & 0.833 & 0.817 & 0.842 & 0.842 & 0.825 & \textbf{0.883} & 0.900 & \textbf{0.908} \\
length-matched summary    & 1.7K  & 0.892 & 0.892 & 0.892 & 0.883 & 0.800 & 0.825 & 0.842 & 0.817 \\
raw trace pool (verbatim) & 40.4K & 0.850 & 0.825 & 0.850 & 0.825 & 0.867 & 0.858 & 0.850 & 0.767 \\
induced taxonomy          & 1.9K  & \textbf{0.900} & \textbf{0.917} & \textbf{0.925} & \textbf{0.917} & 0.825 & 0.833 & 0.858 & 0.842 \\
taxonomy $+$ 3 labeled runs & 4.0K & 0.892 & 0.858 & 0.892 & 0.858 & \textbf{0.875} & 0.850 & \textbf{0.917} & 0.900 \\
\bottomrule
\end{tabular}
\end{table}

\begin{table}[t]
\caption{\textbf{Functional substitution on Terminal-Bench, all pool sizes}
(balanced accuracy, pooled over two pool seeds; decision-token protocol).
Tokens = mean prompt tokens at $N{=}40$. No condition contrast reaches
significance at any $N$ for either consumer: the taxonomy matches the
verbatim pool at ${\sim}15\times$ fewer total tokens.}
\label{tab:functional_tb}
\centering
\small
\begin{tabular}{lccccccccc}
\toprule
 & & \multicolumn{4}{c}{Qwen3.5-27B} & \multicolumn{4}{c}{Qwen3.5-122B-A10B} \\
\cmidrule(lr){3-6}\cmidrule(lr){7-10}
Context & tokens & $N{=}5$ & $N{=}10$ & $N{=}20$ & $N{=}40$ & $N{=}5$ & $N{=}10$ & $N{=}20$ & $N{=}40$ \\
\midrule
no failure context        & 7.1K   & \multicolumn{4}{c}{0.840} & \multicolumn{4}{c}{0.800} \\
3 random traces           & 15.7K  & 0.830 & 0.830 & 0.820 & 0.840 & 0.830 & 0.830 & 0.840 & 0.830 \\
length-matched summary    & 7.8K   & 0.790 & 0.840 & 0.840 & 0.790 & 0.780 & 0.800 & 0.810 & 0.790 \\
raw trace pool (verbatim) & 121.0K & 0.840 & 0.840 & 0.820 & 0.840 & 0.850 & 0.820 & 0.830 & 0.810 \\
induced taxonomy          & 8.3K   & 0.800 & 0.800 & 0.800 & 0.820 & 0.780 & 0.800 & 0.780 & 0.790 \\
taxonomy $+$ 3 labeled runs & 17.1K & 0.840 & 0.850 & 0.810 & 0.810 & 0.830 & 0.840 & 0.830 & 0.830 \\
\bottomrule
\end{tabular}
\end{table}

\section{TRAIL Faithfulness Protocol}
\label{app:trail}

This appendix gives the full protocol behind the TRAIL faithfulness result in \Cref{sec:trail_main}.

\subsection{Dataset and protocol}

TRAIL~\citep{deshpande2025trail} contains 117 GAIA-derived agent traces hand-annotated by four software-engineering experts under a 20-category canonical taxonomy. We use TRAIL only as an external validation set; no TRAIL labels are used to induce the \methodname{} taxonomy. The panel reaches its final label set through span-grounded prompts, one round of peer deliberation, and majority vote.

\subsection{Panel composition}

The panel consists of four LLM annotators: Claude Opus~4.7, GPT-5.4, Gemini-2.5-Pro, and GPT-5.4-mini. Each annotator receives the trace, the induced taxonomy, and span-grounding instructions requiring evidence for every fired code. After independent annotation, the panel performs one round of peer deliberation and outputs a majority-vote label set.

\subsection{Agreement metrics}

We report agreement at the failure-area level because the induced taxonomy and TRAIL's hand-crafted taxonomy do not share identical leaf labels. Area-level mapping groups semantically equivalent leaf codes into broad failure areas (Reasoning / Planning / System) before computing Cohen's $\kappa$ against the expert gold. Bootstrap 95\% confidence intervals use $n_{\text{boot}}{=}2000$ trace-level resamples. The two headline numbers use different consensus thresholds: the matched-panel $0.682$ is the thr$=$1 cell of the cross-family no-grounding row, and the full-protocol $0.725$ is the thr$=$2 cell of the span-grounded$+$deliberation row. The same-model row of Table~\ref{tab:trail_bootstrap} isolates the effect of panel diversity: replacing the cross-family panel with four GPT-5.4 annotators lowers area-$\kappa$ from $0.682$ to $0.625$ at thr$=$1 (the comparison narrows and reverses at higher thresholds, where the CIs overlap).

\begin{table}[t]
\centering
\small
\caption{TRAIL panel-vs-gold area-$\kappa$ under different methodology variants. Each row is the same 117-trace evaluation with a different protocol; thresholds are the number of panel members (out of 4) that must agree on a category. CIs are 95\% percentile intervals from 2000 trace-level bootstrap resamples.}
\label{tab:trail_bootstrap}
\begin{tabular}{lcccccc}
\toprule
Methodology & thr=1 & 95\% CI & thr=2 & 95\% CI & thr=3 & 95\% CI \\
\midrule
Cross-family 4-panel (no grounding) & 0.682 & $[0.608, 0.754]$ & 0.500 & $[0.476, 0.611]$ & 0.307 & $[0.236, 0.380]$ \\
+1-round deliberation & 0.692 & $[0.620, 0.764]$ & 0.590 & $[0.517, 0.665]$ & 0.439 & $[0.354, 0.524]$ \\
Same-model 4$\times$ GPT-5.4 & 0.625 & $[0.544, 0.697]$ & 0.630 & $[0.548, 0.710]$ & 0.563 & $[0.485, 0.640]$ \\
Span-grounded 4-panel & 0.725 & $[0.651, 0.796]$ & 0.563 & $[0.492, 0.635]$ & 0.404 & $[0.335, 0.473]$ \\
\textbf{Span-grounded + deliberation} & \textbf{0.761} & $\mathbf{[0.690, 0.824]}$ & \textbf{0.725} & $\mathbf{[0.654, 0.795]}$ & 0.645 & $[0.573, 0.717]$ \\
\bottomrule
\end{tabular}
\end{table}

\subsection{Vocabulary comparison}

Holding the panel and prompts fixed and varying only the vocabulary, TRAIL's hand-crafted 20-category taxonomy yields area-$\kappa = 0.516$, while \methodname{}'s 29 induced codes yield area-$\kappa = 0.682$. The full \methodname{} protocol reaches area-$\kappa = 0.725$ [95\% CI $0.654$, $0.795$], substantial agreement on the Landis~\&~Koch scale, with the lower CI bound exceeding the conventional $0.61$ threshold.

\begin{table}[h]
\centering
\small
\caption{\textbf{TRAIL vocabulary comparison.} The panel and prompts are held fixed; only the vocabulary the annotators consume changes.}
\label{tab:trail_vocab_appendix}
\begin{tabular}{lc}
\toprule
Vocabulary used by panel & Area-level $\kappa$ vs expert gold \\
\midrule
TRAIL hand-crafted 20-category vocabulary & $0.516$ \\
\methodname{} induced vocabulary & $\mathbf{0.682}$ \\
\methodname{} induced vocabulary, full panel protocol & $\mathbf{0.725}$ \\
\bottomrule
\end{tabular}
\end{table}

\subsection{Leaf-level diagnostics}

Fine-grained leaf-level agreement is lower because the induced taxonomy is not designed to reproduce TRAIL's 20 labels exactly. We therefore report leaf-level results only as a diagnostic. The main validation target is area-level agreement, where semantic failure areas can be compared across different vocabularies. At the 20-leaf level, panel-vs-gold $\kappa$ remains in the fair range ($\kappa \approx 0.34$), consistent with the difficulty of fine-grained failure annotation reported in the published TRAIL paper.

\begin{table}[h]
\centering
\small
\caption{Individual annotator metrics vs TRAIL gold after grounded prompts and one round of deliberation.}
\label{tab:trail_per_annotator}
\begin{tabular}{lcccc}
\toprule
Annotator & leaf-$\kappa$ & area-$\kappa$ & area Jaccard & Landis \& Koch (area) \\
\midrule
Claude Opus 4.7    & 0.334 & 0.744 & 0.821 & substantial \\
GPT-5.4-mini       & 0.308 & 0.676 & 0.783 & substantial \\
GPT-5.4            & 0.317 & 0.625 & 0.729 & substantial \\
Gemini 2.5 Pro     & 0.301 & 0.542 & 0.650 & moderate \\
\bottomrule
\end{tabular}
\end{table}

\begin{table}[h]
\centering
\small
\caption{Pairwise inter-annotator $\kappa$ (leaf-level / area-level) on TRAIL after grounded prompts and deliberation. All six pairs reach $\kappa \geq 0.587$ at leaf level.}
\label{tab:trail_pairwise}
\begin{tabular}{lcccc}
\toprule
 & Opus 4.7 & GPT-5.4 & GPT-5.4-mini & Gemini 2.5 Pro \\
\midrule
Opus 4.7        & --- & 0.742 / 0.791 & 0.794 / 0.814 & 0.604 / 0.688 \\
GPT-5.4         &     & ---           & 0.710 / 0.808 & 0.651 / 0.687 \\
GPT-5.4-mini    &     &               & ---           & 0.587 / 0.671 \\
Gemini 2.5 Pro  &     &               &               & --- \\
\bottomrule
\end{tabular}
\end{table}

\subsection{Mapping induced codes to TRAIL categories}

The judge step that maps each \methodname{}-generated code to its closest TRAIL canonical ID finds a many-to-one mapping: several broad TRAIL categories receive multiple finer-grained induced codes. The mapping is high-precision: 28 of 29 \methodname{} codes (96.6\%) admit a clean TRAIL counterpart.

\begin{table}[h]
\centering
\small
\caption{Auto-to-hand-crafted code mapping. Each \methodname{} code expresses a more specific failure mode than the broader TRAIL category.}
\label{tab:trail_code_mapping}
\begin{tabular}{lp{8.5cm}}
\toprule
TRAIL category (gold) & \methodname{} auto-codes that map to it \\
\midrule
R5 Incorrect Problem Identification & B.1 Tracker\_Incorrect\_Tag\_Configuration; B.2 Tracker\_Incomplete\_Instrumentation\_Coverage; C.1 Entity\_Identity\_Collapse; C.2 Temporal\_Anchor\_Mismatch; C.6 Tabular\_Filter\_Or\_Aggregation\_Misapplication; C.9 Spatial\_Or\_Ordering\_Frame\_Misinterpretation \\
R4 Tool Output Misinterpretation & B.9 Orchestrator\_Evidence\_Integration\_Failure; C.5 Retrieved\_Fact\_To\_Answer\_Transformation\_Error; C.7 Exact\_Span\_Extraction\_Error; C.8 Multimedia\_Perceptual\_Misread; C.10 Cross\_Source\_Schema\_Mismatch; C.13 Cross\_Source\_Synthesis\_Reconciliation \\
P4 Task Orchestration Errors & A.3 Strategy\_Initialization\_Or\_Routing\_Failure; B.4 Unknown\_Role\_Misalignment; B.8 Orchestrator\_Incorrect\_Tool\_Selection; B.10 Orchestrator\_Insufficient\_Self\_Verification; C.3 Incomplete\_Multihop\_Chain \\
R8 Instruction Non-compliance & A.1 Final\_output\_format\_noncompliance; B.3 Tracker\_Unsafe\_Or\_Noncompliant\_Tracking\_Method; C.12 Required\_Source\_Type\_Not\_Actually\_Used \\
P1 Context Handling Failures & A.2 Self\_inconsistency\_across\_pipeline\_states; A.6 Interstage\_Handoff\_Loss\_Or\_Drift \\
\bottomrule
\end{tabular}
\end{table}

\section{OlympiadBench Mechanism Details}
\label{app:olympiad_mechanism}

The main text uses OlympiadBench as the mechanism study for search because it has the largest held-out set and the longest recorded evolution trace. This appendix contains the supporting artifacts: the raw score trajectory, judge outputs around the breakthroughs, non-breakthrough base rates, the failure-burden over evolution, the four-arm failure profile, and architecture diffs.

\subsection{Raw score trajectory}
\label{app:olympiad_score_curves}

\begin{figure}[h]
\centering
\includegraphics[width=0.85\linewidth]{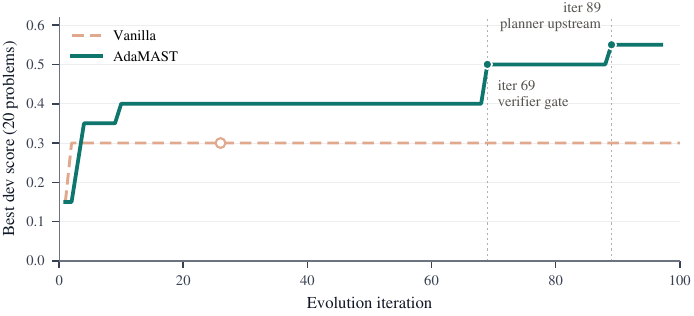}
\caption{\textbf{OlympiadBench best-dev-score trajectory.} Per-iteration running max of dev score (20 problems) for the vanilla run (search ends at iteration 26, max 0.30; the dashed line is continued flat for comparison) and the \methodname{} run (97 iterations, teal). Markers annotate the iter-69 verifier-gate breakthrough and the iter-89 planner-upstream breakthrough.}
\label{fig:score_curves_olympiad}
\end{figure}

\Cref{fig:score_curves_olympiad} is the raw best-dev-score-per-iteration curve for the OlympiadBench \methodname{} and vanilla runs, with the two annotated score-jumps at iter~69 (verifier gate) and iter~89 (planner upstream). The vanilla run saturates at iter~26 with max dev score $0.30$; \methodname{} crosses to $0.40$ at iter~10, $0.50$ at iter~69, and $0.55$ at iter~89.

\subsection{Breakthrough judge outputs}
\label{app:breakthrough_judge}

For each score jump (iteration 67$\to$69 and iteration 79$\to$89), we recover the parent iteration, the child iteration, the fired codes in the parent's \texttt{taxonomy\_judge} artifact, and the architecture edit introduced in the child. The verbatim judge outputs below show that the code-driven mutations were grounded in concrete trace evidence.

\paragraph{Iteration-43 judge output (on the lineage of the iter-69 breakthrough, dev $0.40 \to 0.50$).}
\begin{small}
\begin{verbatim}
A-codes:
  [A.1] Behavioral Anomaly: Non-convergent iteration / stagnation
    Evidence: "Output: FINAL ANSWER: Please provide the problem statement.";
              "Output: FINAL ANSWER: 0";
              "Output: FINAL ANSWER: Cannot determine without the problem statement."
  [A.3] Instruction Compliance Violation
    Evidence: "Output: FINAL ANSWER: Please provide the problem statement.";
              "OUTPUT: FINAL ANSWER: Problem statement missing."
  [A.6] Ground-Truth / Verifier Mismatch
    Evidence: Coordinator Aggregation Mismatch notes;
              inconsistent final answers across branches.

B-codes (by role):
  checker:
    [B.1] Non-compliant Output by Checker -- checkers default to problem-
          statement prompts instead of solving.
  refiner:
    [B.3] Refiner Fails to Improve or Regresses After Critique
    [B.4] Refiner Overcorrects or Hijacks Methodology
  coordinator:
    [B.6] Coordinator Aggregation Mismatch: aggregated results conflict
          due to divergent FINAL ANSWER lines; no single coherent vector.

C-codes:
  [C.2] Edge Case Omission / Degeneration
  [C.3] Overgeneralization in Domain Prereqs
  [C.4] Algebraic Manipulation Gap
  [C.5] Combinatorics Counting Misreasoning
\end{verbatim}
\end{small}

\paragraph{Iteration-60 judge output (on the lineage of the iter-89 breakthrough, dev $0.50 \to 0.55$).}
\begin{small}
\begin{verbatim}
A-codes:
  [A.3] Instruction Compliance Violation: Output formats and content
        frequently violate task prompts.
  [A.6] Ground-Truth / Verifier Mismatch: Divergent final answers across
        checkers for the same problem.
  [A.4] Evaluation Feedback Misinterpretation: Critique echoed but not
        corrected.

B-codes:
  refiner:
    [B.3] Refiner Fails to Improve or Regresses After Critique.
    [B.4] Refiner Overcorrects or Hijacks Methodology.

C-codes:
  [C.2] Edge Case Omission / Degeneration.
  [C.3] Overgeneralization in Domain Prereqs.
  [C.4] Algebraic Manipulation Gap.
  [C.5] Combinatorics Counting Misreasoning.
\end{verbatim}
\end{small}

\subsection{Non-breakthrough base rates}

We compare code firings on breakthrough parents against non-breakthrough parents. Codes that fire frequently everywhere are not sufficient explanations of score gains; codes that are enriched on breakthrough parents are more plausible mutation signals. Across the 97-iteration run, the codes that fire as primary on breakthrough parents (A.3, B.3, C.2 for the iter-69 breakthrough; A.6, B.6, C.3 for the iter-89 breakthrough) are present in the seed taxonomy or bootstrap, indicating that the breakthroughs are driven by the mutator acting on existing taxonomy entries rather than by newly-introduced refinement codes.

\begin{figure}[h]
\centering
\includegraphics[width=\linewidth]{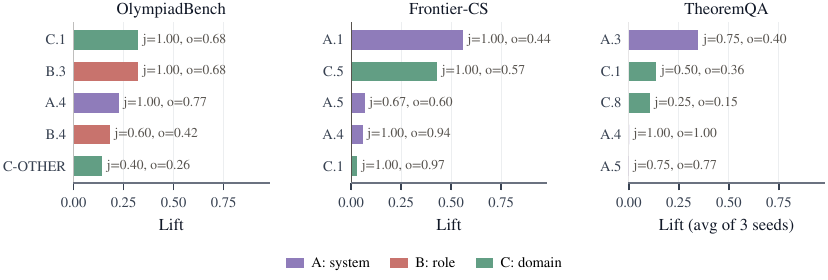}
\caption{\textbf{Per-code lift on score jumps across three benchmarks.} For each benchmark we identify the iterations where dev score jumps and compute each code's lift = $P(\text{fire} \mid \text{jump}) - P(\text{fire} \mid \text{non-jump})$. Bar annotations give the two firing rates directly: $j = P(\text{fire} \mid \text{jump})$, the fraction of jump iterations where the code fired, and $o = P(\text{fire} \mid \text{non-jump})$, the same fraction over all other iterations. Jump bases per panel: OlympiadBench 5 jump vs.\ 31 non-jump iterations (28 codes); Frontier-CS 3 vs.\ 70 (14 codes); TheoremQA 3 vs.\ 59 totaled over three seeds (14--15 codes per seed). The top codes per benchmark line up with the role- and domain-specific codes named in the verbatim judge outputs.}
\label{fig:gain_attribution}
\end{figure}

\Cref{fig:gain_attribution} extends this analysis to per-code \emph{lift} on score-jump iterations across three benchmarks. For each benchmark we score every code by its firing rate on the score-jump iterations ($j$) and on the remaining iterations ($o$), then plot the top few by lift ($j-o$). The bench-level top codes are the role- and domain-specific ones identified in the verbatim judge outputs above, not generic A.1/A.5 codes; this is the cross-benchmark version of the OlympiadBench mechanism argument.

\subsection{Failure burden over evolution}
\label{app:olympiad_axis_shift}

The \methodname{} run retires 12 of 28 codes by iteration 92. Severity-weighted failure burden decreases by 23\%, from 16.9 to 13.0, even though the raw firing count stays roughly constant. The B (role-specific) share grows from $39.8\%$ in iterations 1--30 to $46.9\%$ in iterations 61--97, while the C (domain) share contracts from $32.0\%$ to $25.5\%$. Each of the two score breakthroughs (iterations 69 and 89) follows a taxonomy-refinement event (iterations 29 and 78, respectively); a final refinement at iteration 92 closes the run.

\begin{figure}[h]
\centering
\includegraphics[width=0.85\linewidth]{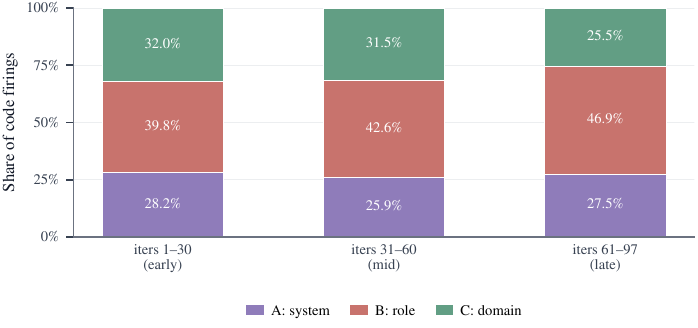}
\caption{\textbf{Failure-axis distribution shifts during evolution.} Share of judge-fired codes by axis in three windows of the OlympiadBench \methodname{} run (early: iterations 1--30; mid: 31--60; late: 61--97). The B-axis share grows over the run as the architecture acquires more roles; the C-axis share contracts as the solver acquires domain-specific guard rails.}
\label{fig:axis_shift}
\end{figure}

\begin{figure}[h]
\centering
\includegraphics[width=0.85\linewidth]{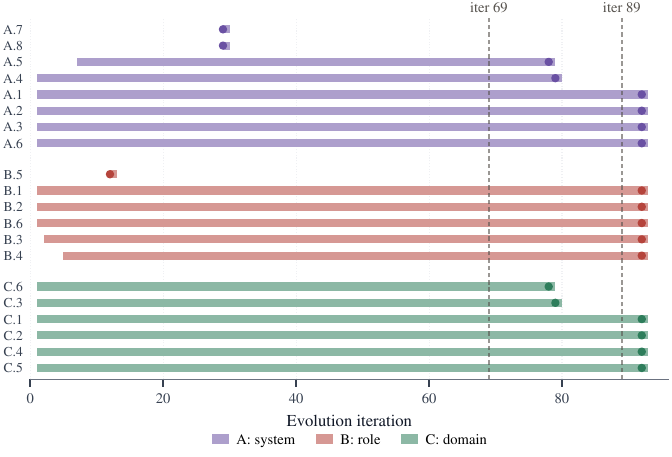}
\caption{\textbf{Code lifecycle on the OlympiadBench \methodname{} run.} Each bar shows the iteration range over which a code fired at least once; the dot marks the last-firing iteration. Of the codes here, seven retire fully before iteration~80; the rest persist through the iteration-92 final taxonomy snapshot.}
\label{fig:code_lifecycle}
\end{figure}

\begin{figure}[h]
\centering
\includegraphics[width=0.85\linewidth]{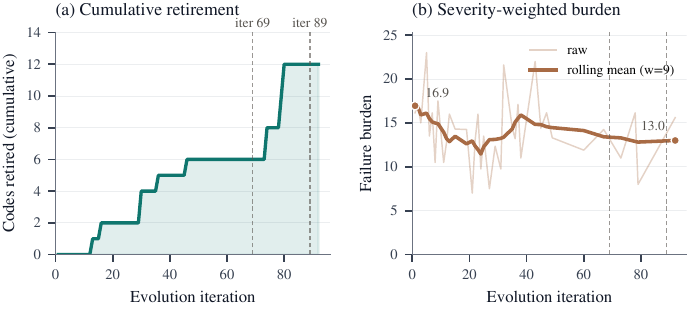}
\caption{\textbf{Failure-mode retirement and burden decay over evolution.} \textbf{(a)} Cumulative count of retired failure codes (codes that never fire again after a given iteration); 12 of the 28 codes have stopped firing by iteration~92, with the two score breakthroughs (dashed lines, iterations 69 and 89) marked. \textbf{(b)} Severity-weighted failure burden: the aggregate burden falls by 23\% (16.9 to 13.0) from iteration 1 to iteration 92 as failure modes retire and the residual concentrates on harder failures.}
\label{fig:failure_decay}
\end{figure}

\begin{figure}[h]
\centering
\includegraphics[width=\linewidth]{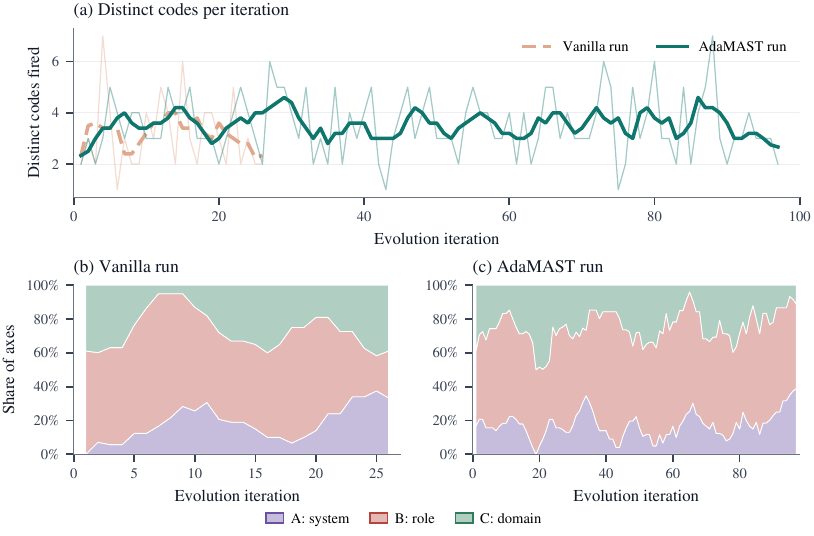}
\caption{\textbf{Vanilla vs \methodname{} per-iteration failure surface.} \textbf{(a)} Distinct codes fired per iteration. \textbf{(b)} Stacked share of A/B/C axes for the vanilla run measured post-hoc against the final \methodname{} taxonomy (26 iters). \textbf{(c)} Same stacked share for the \methodname{} run (97 iters). Adaptive feedback does not produce more labels per iteration; it reshapes which axes the failures concentrate on.}
\label{fig:notax_vs_adamast}
\end{figure}

\Cref{fig:notax_vs_adamast} compares the per-iteration code-firing surface of the vanilla and \methodname{} runs directly. Panel~(a) shows that both runs carry a similar number of distinct fired codes per iteration once \methodname{} has its bootstrap codes; the difference is not ``more labels''. Panels~(b) and~(c) show how the share between A/B/C axes evolves: the vanilla run can only be \emph{measured} against the post-hoc taxonomy, so its axis-share fluctuates from iteration to iteration without any directional shift; the \methodname{} run shows the B-share grow steadily as the architecture acquires more roles.

\subsection{Post-hoc four-arm failure profile}

We apply the final OlympiadBench taxonomy as a fixed measurement instrument to the seed, vanilla, MAST-guided, and \methodname{}-guided final architectures. This produces a comparable failure profile for the four arms on the 655-problem held-out set.

\begin{table}[h]
\centering
\small
\caption{Primary failure-code firing counts per (arm, code) on OlympiadBench, $n{=}655$ problems per arm. Per-arm totals: seed 107, vanilla 79, MAST 61, \methodname{} 55.}
\label{tab:olympiad_histogram_counts}
\begin{tabular}{llrrrr}
\toprule
Code & Name & Seed & Vanilla & MAST & \methodname{} \\
\midrule
A.1   & \emph{Output format violation}                & 16 & 17 & 7  & 0  \\
A.2   & \emph{Premature pipeline termination}         & 3  & 1  & 4  & 2  \\
A.8   & \emph{Instruction non-compliance}             & 1  & 3  & 5  & 0  \\
A.10  & \emph{Aggregation/merge error}                & 2  & 2  & 4  & 1  \\
B.3   & \emph{Checker final answer incorrect}         & 0  & 0  & 0  & 8  \\
B.4   & \emph{Solver low-quality reasoning}           & 67 & 42 & 32 & 34 \\
C.6   & \emph{Incomplete reasoning / missing lemmas}  & 3  & 1  & 3  & 0  \\
C.7   & \emph{Misinterpretation of definition}        & 1  & 8  & 4  & 6  \\
C.12  & \emph{Combinatorial overcount / symmetry}     & 7  & 0  & 0  & 0  \\
\bottomrule
\end{tabular}
\end{table}

\begin{figure}[h]
\centering
\includegraphics[width=0.95\linewidth]{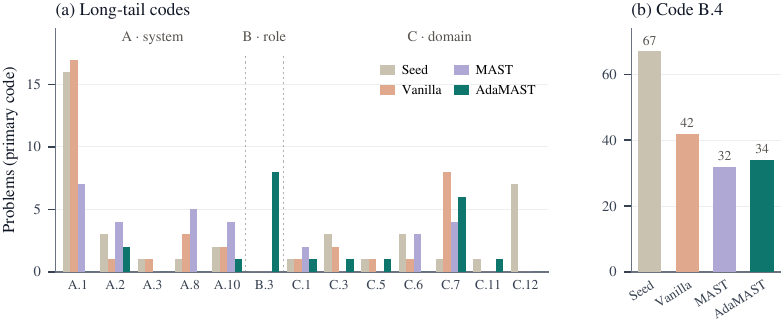}
\caption{\textbf{Cross-arm failure profile on OlympiadBench.} The post-hoc judge assigns the primary failure-mode label to each (arm, instance) cell on the 655-problem held-out set using the iteration-92 taxonomy.}
\label{fig:olympiad_histogram}
\end{figure}

\subsection{Architecture diffs}

The \methodname{}-guided best architecture (iter~92, test 91.91\%) runs four diversified solver variants under a 25-call LLM budget, applies a lightweight verifier that boosts the score of structurally-valid candidates (\texttt{c["score"] += 0.1} on \texttt{VALID}), selects up to four diverse candidates for refinement via a length-and-variant-keyed frontier, and runs up to two refinement rounds before final aggregation. The verifier replaces the binary gate of earlier iterations with a numerical score boost, the iter-92 response to the B.6 \textit{Coordinator Aggregation Mismatch} code raised by the iteration-60 judge.

The vanilla best architecture (test 87.94\%) is an adversarial-debate pipeline: a primary solver, two critics, an improver consolidating critiques into a revision, two more critics, a second improver, and a synthesis step. The MAST-guided best architecture (test 89.47\%) is a strict five-stage pipeline: solver, refiner, voter, finalizer, terminal output generator.

The mutator was capable of producing any of these structures from the same seed. \methodname{} discovered a verifier-gated four-variant architecture because the categorical failure codes (B.6 \textit{Coordinator Aggregation Mismatch}, B.4 \textit{Refiner Overcorrects}, C.5 \textit{Combinatorics Counting Misreasoning}) pointed the mutator at parallel diversification; the vanilla mutation prompts contained only free-text reflection on the previous candidate's score.

\section{Runtime Monitoring Details}
\label{app:runtime_details}

This appendix gives details for the runtime monitoring experiments in \Cref{sec:runtime_results}: the SWE-agent checkpoint protocol, the Claude Code runtime skill setup, and the blind-judge contamination-scrubbing protocol. We detail SWE-agent first because its arm ladder includes a free-text reflection rung (Reflexion), separating the value of checkpointing itself from the value of categorical anchoring.

\subsection{SWE-agent checkpoint protocol}
\label{app:sweagent_runtime}

Each SWE-agent run uses the same GPT-5 solver and the same task instances; the arms differ only in what anchors the checkpoints. Base receives no reflection of any kind. Reflexion adds prompted free-text reflection at the checkpoints; MAST in-prompt anchors the same checkpoints to the published 14-code MAST taxonomy; \methodname{} in-prompt anchors them to the induced SWE-bench taxonomy. All three reflection arms share the identical in-prompt checkpoint scaffold.

Checkpoints fire at three points: after initial repository exploration, after the first round of edits, and before final submission. SWE-agent has no native hook mechanism, so compliance is enforced by a submit-gate inside \texttt{tools/review\_on\_submit\_m}, which validates that three checkpoint markers are present before allowing submission. At each checkpoint the agent produces a structured summary of recent actions, the current diff, recent tool output, and unresolved hypotheses.
\subsection{Main SWE-agent result}

The headline four-arm numbers appear in \Cref{tab:swebench_sweagent} of the main text: Base 25/50, Reflexion 30/50, MAST in-prompt 34/50, and \methodname{} in-prompt 35/50. The ordering tracks the structure of the feedback: anchoring the checkpoint self-check to a categorical vocabulary accounts for most of the gain over unanchored free-text critique, with the induced, harness-specific vocabulary adding a further margin over the fixed MAST checklist.

\begin{figure}[h]
\centering
\includegraphics[width=\linewidth]{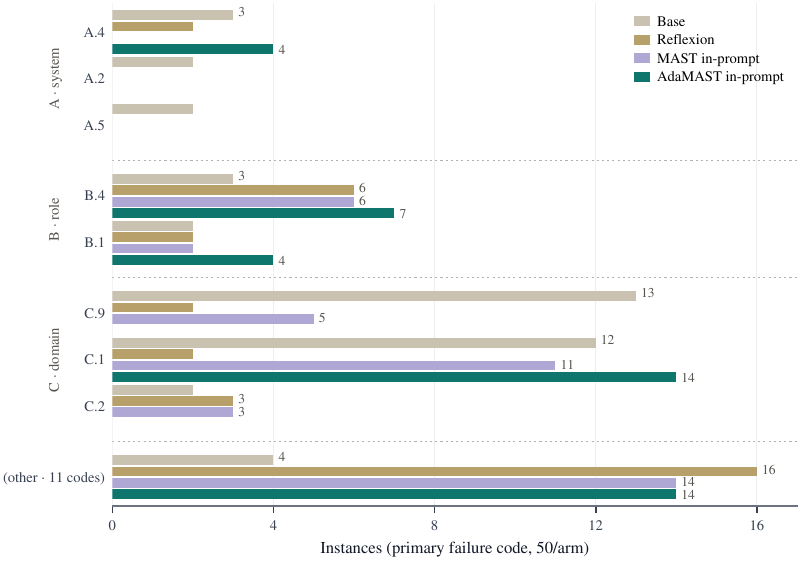}
\caption{\textbf{Primary failure code per (arm, instance) cell under the SWE-agent harness.} Two codes dominate Base's failures: C.9 \emph{unclean patch contents} (13 firings) and C.1 \emph{ad-hoc validation only} (12 firings). \methodname{} in-prompt retires C.9 entirely (13 to 0) while C.1 persists at a similar level (14 vs.\ Base's 12): the induced vocabulary eliminates the patch-hygiene failure but shallow validation remains the dominant residual mode.}
\label{fig:swebench_sweagent_histogram}
\end{figure}

\begin{figure}[h]
\centering
\includegraphics[width=0.80\linewidth]{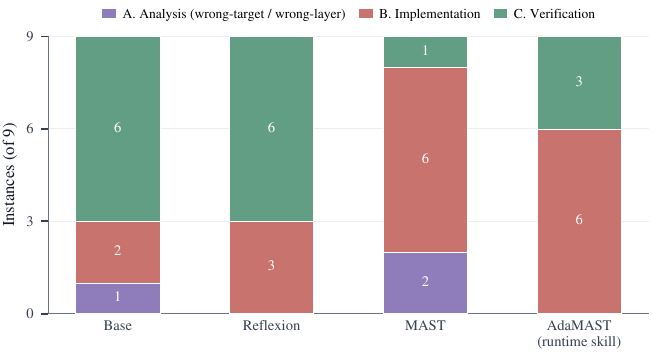}
\caption{\textbf{SWE-bench failure-area composition per arm.} Each bar is the 9 instances unresolved by every arm under that condition, decomposed into three failure areas. The composition shifts upstream as the feedback surface gets richer: Verification dominates under Base/Reflexion $\rightarrow$ Implementation dominates under MAST and \methodname{} arms.}
\label{fig:swe_failure_categories}
\end{figure}

\Cref{fig:swe_failure_categories} aggregates the same instance-level labels into the three high-level failure areas (Analysis: wrong-target / wrong-layer fixes; Implementation; Verification). The Base and Reflexion arms fail mostly on Verification (6 of 9 instances each; the agent submits without verifying its recent actions); under MAST and the \methodname{} arms, the Verification share drops (to 1 and 3 of 9, respectively) and the residual failures concentrate in Implementation (6 of 9 each).

\subsection{Secondary Claude Code harness}
\label{app:claudecode_secondary}

We also tested the same taxonomy feedback in a Claude Code harness, delivered here through the harness's native skill mechanism (Haiku 4.5 solver, three arms, 50 Verified Mini instances); the three-seed headline results appear in \Cref{tab:swebench_cc} of the main text. Claude Code includes stronger native self-correction through direct test execution, which changes what the feedback layer contributes. Categorical taxonomies still improve over the base agent; in the seed-1 run analyzed in the remainder of this appendix, the per-arm counts are 30/50 (Base), 34/50 (MAST runtime skill), and 35/50 (\methodname{} runtime skill). 

\begin{figure}[h]
\centering
\includegraphics[width=\linewidth]{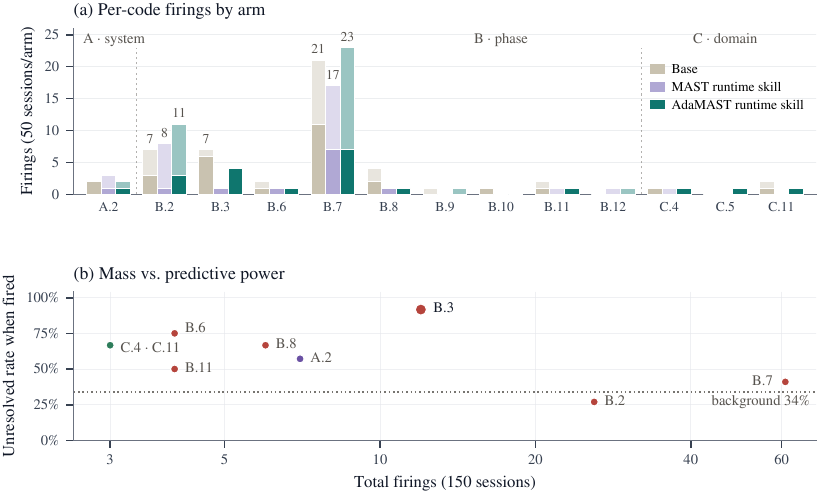}
\caption{\textbf{Per-arm failure code firings under the Claude Code harness.} Single-seed run on 50 Verified Mini instances, three arms (Base, MAST runtime skill, and \methodname{} runtime skill). \textbf{(a)} Per-code firings by arm; within each bar, the solid segment counts firings on unresolved sessions and the light segment firings on resolved sessions. \textbf{(b)} Mass vs.\ predictive power: per-code total firings across the 150 sessions against the unresolved rate when the code fires, with the 34\% background unresolved rate marked; B.3 is rare but strongly predictive of failure, while B.7 is common but near background.}
\label{fig:swebench_cc_histogram}
\end{figure}

\paragraph{Harness and solver.}
\label{app:claudecode_setup}
All three arms of the Claude Code study use Claude Code (the Anthropic-released agent CLI) with the Haiku 4.5 solver (\texttt{claude-haiku-4-5}) under a 1800-second wall-clock timeout and the harness's default token limits. The solver runs natively against a per-instance Docker container from the SWE-bench evaluation image set (\texttt{swebench/sweb.eval.x86\_64.django\_*} and \texttt{swebench/sweb.eval.x86\_64.sphinx-doc\_*}); the workspace is volume-mounted at \texttt{/testbed} and the agent interacts with the container through a host-side wrapper that issues \texttt{docker exec} into the \texttt{conda activate testbed} environment. This lets the agent run \texttt{pytest} between edits and observe pass/fail within its own loop, while isolating each session's environment.

\paragraph{Three arms.}
The arms vary only the failure-feedback layer around the same Haiku 4.5 solver on the same 50 instances. \textbf{Base} runs vanilla Claude Code. \textbf{MAST runtime skill} delivers the published 14-code MAST taxonomy~\citep{cemri2025mast} as a project-local skill loaded into the agent's context. \textbf{\methodname{} runtime skill} delivers the 32-code Claude Code SWE-bench taxonomy (\Cref{tab:swe_cc_codes_a,tab:swe_cc_codes_b,tab:swe_cc_codes_c}) through the identical integration. Each arm's skill files live in a per-arm project-local directory; cross-arm probes verified zero residual references to other arms' files in any session trajectory.

\paragraph{Checkpoint delivery.} The MAST and \methodname{} arms register checkpoints through the harness's native hooks, which fire after tool calls, on sub-agent completion, and at task completion. A checkpoint that surfaces fired codes triggers the bounded repair loop of \Cref{sec:runtime_method}: at most three attempts, after which the agent reports remaining issues rather than claiming success.

\begin{table}[h]
\caption{Per-arm cost and median per-session wall-clock duration on the 50-instance SWE-bench Verified Mini under the Claude Code harness. Solver is Haiku 4.5.}
\label{tab:swe_cost_cc}
\centering
\small
\begin{tabular}{lrrr}
\toprule
Arm & Total cost & Mean per instance & Median duration \\
\midrule
Arm 1: Base                    & \$26.38 & \$0.53 & 434\,s \\
Arm 2: MAST runtime skill         & \$24.58 & \$0.49 & 434\,s \\
Arm 3: \methodname{} runtime skill & \$27.24 & \$0.54 & 470\,s \\
\bottomrule
\end{tabular}
\end{table}

\paragraph{32-code Claude Code SWE-bench taxonomy.}
\label{app:swe_taxonomy_claudecode}
The vocabulary the blind post-hoc judge applies is induced from SWE-bench Verified trajectories produced under the Claude Code harness (Base arm only, so the vocabulary is not contaminated by knowledge of which codes the feedback arms would surface). It contains 32 codes total (4~A, 12~B, 16~C). The B-axis is reframed from multi-agent roles into single-agent phases (Edit / Plan / Verify) because Claude Code operates as a single agent with phased behavior rather than distinct solver/coordinator/checker roles.

\begin{table}[h]
\centering
\small
\setlength{\tabcolsep}{4pt}
\caption{System-level (A) codes that fire as concerns in the SWE-bench taxonomy (Claude Code harness). Frequencies are over the 150 judged sessions of the three arms. Three further A-codes are defined in the taxonomy (A.1 \emph{Output format noncompliance}, A.3 \emph{Runaway repetition or looping}, A.4 \emph{Unwarranted refusal or abandonment}) but did not fire on any judged session.}
\label{tab:swe_cc_codes_a}
\begin{tabular}{llp{5cm}r}
\toprule
Code & Name & Description & Freq. \\
\midrule
A.2 & Execution environment mismatch & Runtime environment differs from assumptions (missing tools/dependencies, version skew) & 4.7\% \\
\bottomrule
\end{tabular}
\end{table}

\begin{table}[h]
\centering
\small
\setlength{\tabcolsep}{4pt}
\caption{Top phase-specific (B) codes in the SWE-bench taxonomy (Claude Code harness). Frequencies are over the 150 judged sessions of the three arms. Phase prefixes \emph{Edit}, \emph{Plan}, \emph{Verify} replace the multi-agent role prefixes used in the SWE-agent taxonomy.}
\label{tab:swe_cc_codes_b}
\begin{tabular}{llp{5cm}r}
\toprule
Code & Name & Description & Freq. \\
\midrule
B.7 & Verify partial run misreported as success & Only a subset of tests is executed and passing reported as global success & 40.7\% \\
B.2 & Edit patch lacks test updates & Functional change without adding or updating relevant tests & 17.3\% \\
B.3 & Edit overbroad patch footprint & Patch changes unrelated files or performs refactors not required for the fix & 8.0\% \\
B.8 & Verify ignored import or syntax errors & Import/syntax errors in logs are downplayed or omitted in the report & 4.0\% \\
B.6 & Plan skips verification loop & Workflow omits running targeted tests after edits & 2.7\% \\
B.11 & Edit proceeds despite tool or file errors & Agent continues with conclusions or success claims after observable tool errors & 2.7\% \\
B.9 & Verify misparsed test summary & Misinterprets test-runner output, leading to incorrect pass/fail conclusion & 1.3\% \\
B.12 & Plan retrieval pattern misconfiguration & Search/grep patterns chosen such that relevant files are missed or noise is over-included & 1.3\% \\
B.10 & Edit claims changes without execution & Agent asserts code was modified or tests passed without performing the corresponding tool-driven actions & 0.7\% \\
B.1 & Edit wrong target module & Patch modifies unrelated files or the wrong component & 0.7\% \\
\bottomrule
\end{tabular}
\end{table}

\begin{table}[h]
\centering
\small
\setlength{\tabcolsep}{4pt}
\caption{Top domain-reasoning (C) codes in the SWE-bench taxonomy (Claude Code harness). Frequencies are over the 150 judged sessions of the three arms. The codes reflect the django/sphinx mix of the Verified Mini instances. The remaining C-codes in the taxonomy address finer-grained domain patterns and are released alongside the run artifacts.}
\label{tab:swe_cc_codes_c}
\begin{tabular}{llp{5cm}r}
\toprule
Code & Name & Description & Freq. \\
\midrule
C.11 & User override ignored & Framework code overwrites user-specified kwargs or configuration & 2.0\% \\
C.4 & Wrong component or layer targeted & Fix implemented in wrong subsystem or abstraction layer & 2.0\% \\
C.5 & Risky monkeypatch of external internals & Relies on monkeypatching third-party internals rather than documented extension points & 0.7\% \\
C.15 & Parser grammar feature omission & Parser fails to recognize a valid language feature & 0.7\% \\
C.7 & Pipeline reordering without interaction analysis & Fix depends on reordering processing steps without analyzing interactions & 0.7\% \\
\bottomrule
\end{tabular}
\end{table}

\paragraph{Blind-judge contamination scrubbing.}
\label{app:claudecode_judge_protocol}
The three arms differ in what taxonomy information enters the agent's context. Base sees nothing; the MAST runtime skill arm sees the 14-code MAST vocabulary; the \methodname{} runtime skill arm sees the Claude Code SWE-bench taxonomy—each loaded into its context via the project-local skill. Judging the raw trajectories under the same taxonomy would feed the post-hoc judge its own vocabulary. We strip taxonomy-revealing content event-by-event before judging.

\emph{Per-arm event-level scrubbing.} For the \methodname{} runtime skill arm, residual mentions of the Claude Code taxonomy's code identifiers (\texttt{A.1}, \ldots, \texttt{C.16}) or code names are scrubbed from text content. For the MAST runtime skill arm, the same scrubbing is applied to MAST's 14 code identifiers and names. Base sees no taxonomy content and requires no scrubbing. Test-wrapper invocations are kept: those are the agent's interaction with the test environment, which is identical across arms and which the judge needs to see. Assistant reasoning blocks are excluded from judge input across all arms; the judge grades artifacts only.

\emph{Validation.} After scrubbing, (i)~zero residual taxonomy tokens appear in any of the 150 judged trajectories; (ii)~mean cleaned-trajectory size is comparable across arms (within 93--100\,K characters), confirming no arm is artificially shortened.

\paragraph{Cross-arm instance structure.}
The 50 SWE-bench Verified Mini instances are partitioned into three regimes across the 450 Claude Code sessions (3 arms $\times$ 50 instances per seed $\times$ 3 seeds). At least 22 instances are resolved by all three arms across all three seeds (the common solvable set, where any reasonable agent succeeds). At least 5 instances are resolved by no arm under any seed (\texttt{django-11885}, \texttt{django-11964}, \texttt{sphinx-11510}, \texttt{sphinx-9229}, \texttt{sphinx-9461}); this is the ceiling for this harness on this benchmark. The remaining instances sit in the discrimination zone where arms differ. \methodname{} expands the reliably-resolved set within this zone: the \methodname{} runtime skill has 28 instances resolved in all three seeds, against Base's 27. Critically, no instance is resolved by \methodname{} in all three seeds while resolved by Base in none: the \methodname{} gain is concentrated in instances that were already occasionally solvable under Base but became reliably so under \methodname{}'s structured feedback. This is the per-instance shape of the \Cref{tab:swebench_cc} result: \methodname{} does not unlock fundamentally new instances on this harness; it converts flaky resolves into reliable ones.

\emph{Judge configuration.} GPT-5, single call per session at temperature 0. Input is the cleaned event stream plus the candidate patch. Output is a JSON object with \texttt{verdict}~$\in$~\{\texttt{submit}, \texttt{revise}\} and a list of \texttt{concerns}, each with a code, axis, and rationale. The 150 judge calls cost under \$5 in API spend.

\paragraph{Per-(arm, code) primary firing counts.}
\label{app:claudecode_per_arm_counts}

\begin{table}[h]
\centering
\small
\caption{Per-(arm, code) firing counts on SWE-bench Verified Mini under the Claude Code harness, $n{=}50$ sessions per arm. Single-seed run.}
\label{tab:swebench_cc_histogram_counts}
\begin{tabular}{lrrr}
\toprule
Code & Base & MAST & \methodname{} runtime skill\\
\midrule
A.2  &  2 &  3 &  2 \\
B.1  &  1 &  0 &  0 \\
B.2  &  7 &  8 & 11 \\
B.3  &  7 &  1 &  4 \\
B.6  &  2 &  1 &  1 \\
B.7  & 21 & 17 & 23 \\
B.8  &  4 &  1 &  1 \\
B.9  &  1 &  0 &  1 \\
B.10 &  1 &  0 &  0 \\
B.11 &  2 &  1 &  1 \\
B.12 &  0 &  1 &  1 \\
C.4  &  1 &  1 &  1 \\
C.5  &  0 &  0 &  1 \\
C.7  &  0 &  1 &  0 \\
C.11 &  2 &  0 &  1 \\
C.15 &  0 &  1 &  0 \\
\midrule
\textbf{Total firings} & \textbf{51} & \textbf{36} & \textbf{48} \\
\textbf{Resolved (out of 50)} & \textbf{30} & \textbf{34} & \textbf{35} \\
\bottomrule
\end{tabular}
\end{table}
\emph{Resolved/unresolved stratification.} Across the 150 sessions, $30+34+35=99$ are resolved and 51 are unresolved, giving a $51/150 = 34\%$ background unresolved rate. Two codes carry the headline observations. \textbf{B.7} \emph{Verify partial run misreported as success}: 61 total firings, unresolved-rate-when-fired $25/61 = 41\%$ vs.\ $34\%$ background (a mass head but a weak predictor). \textbf{B.3} \emph{Edit overbroad patch footprint}: 12 total firings, unresolved-rate-when-fired $11/12 = 92\%$ (a gain head and a strong predictor).

\subsection{Cross-domain runtime evaluation on OfficeQA Pro}
\label{app:officeqa_runtime}

We additionally evaluate the runtime integration on OfficeQA Pro, a document-grounded quantitative-reasoning benchmark outside the software-engineering domain used in the main runtime experiments. We use all 133 hard questions under the oracle-parsed condition, in which the relevant parsed source documents are available to the agent. Both arms use the same Claude Code harness, Claude Haiku~4.5 solver, system prompt, task files, reflection and repair gate, and official OfficeQA scoring procedure. The only difference is whether the gate is anchored to the induced OfficeQA taxonomy.

The taxonomy was induced from 50 OfficeQA baseline transcripts. Model-based consolidation reduced the initial 26-code inventory to 15 codes; no codes or trace labels were manually authored. The resulting taxonomy was frozen before the evaluation run and remained unchanged across all 133 tasks.

\begin{table}[h]
\centering
\small
\caption{\textbf{Cross-domain runtime evaluation on OfficeQA Pro.} Accuracy under the official scorer at four allowable absolute-relative-error thresholds. Each arm is a single run over the same 133 questions.}
\label{tab:officeqa_runtime}
\begin{tabular}{lcccc}
\toprule
Runtime condition & 0\% & 0.1\% & 1\% & 5\% \\
\midrule
Matched gate, no taxonomy
    & 44.4\% (59/133) & 46.6\% & 54.1\% & 63.2\% \\
Matched gate + \methodname{}
    & \textbf{51.9\% (69/133)} & \textbf{53.4\%} & \textbf{60.2\%} & \textbf{63.9\%} \\
\bottomrule
\end{tabular}
\end{table}

At the strict 0\% threshold, \methodname{} resolves 69 tasks compared with 59 for the matched gate-only control, an improvement of 7.5 percentage points. Of the paired outcomes, 20 tasks are resolved only by \methodname{}, 10 only by the control, 49 by both, and 54 by neither. The corresponding two-sided exact McNemar test gives $p \approx 0.10$. The advantage remains positive across all reported numerical-tolerance bands, although it narrows at the most permissive threshold.

Because this experiment contains one run per arm, we present it as a cross-domain demonstration of the runtime integration rather than a standalone estimate of its average effect. Together with the SWE-bench experiments, it shows that the same taxonomy-conditioned checkpoint mechanism can be applied beyond software engineering to document-grounded quantitative reasoning.

\section{Extended Design Ablations}
\label{app:extended_design_ablations}

\subsection{Injection-location ablation}
\label{app:injection_location}

We ask where taxonomy feedback should enter the mutation prompt. Holding the taxonomy fixed, appending the diagnosis to the evaluation results performs best.

\begin{table}[h]
\centering
\small
\caption{\textbf{Injection-location ablation on TheoremQA.} The same taxonomy feedback is inserted at three locations in the mutation prompt; single-seed (seed 42), reported as directional like the other small ablations. Appending it to evaluation results performs best.}
\label{tab:injection_location}
\begin{tabular}{lc}
\toprule
Injection position & Accuracy \\
\midrule
Appended to evaluation results & \textbf{63.3\%} \\
Inline comments in code & 53.3\% \\
System message & 46.7\% \\
\bottomrule
\end{tabular}
\end{table}

Appending feedback to evaluation results works best because the mutation LLM processes it in the same context as its assessment of program quality; system-message placement is treated as background instruction.

\subsection{Feedback-form diagnostic}
\label{app:feedback_form_diagnostic}

We compare three feedback presentations on TheoremQA: structured \methodname{} codes, a prose summary of the same judge output, and raw trace excerpts. The underlying judge-trace content is held fixed.

\begin{table}[h]
\centering
\small
\caption{\textbf{Feedback-form diagnostic on TheoremQA.} Mean post-search accuracy across three seeds.}
\label{tab:feedback_form}
\begin{tabular}{lcc}
\toprule
Feedback form & Mean post-search accuracy & Per-seed scores \\
\midrule
Structured \methodname{} codes & $56.7\%$ & $56.7,60.0,53.3$ \\
Prose summary & $61.1\%$ & $56.7,70.0,56.7$ \\
Raw trace excerpts & $56.7\%$ & $50.0,53.3,66.7$ \\
\bottomrule
\end{tabular}
\end{table}

The three presentations fall within seed noise of one another (the prose mean is carried by a single high seed), and raw excerpts do not lag far behind. This is the format-null result referenced in \Cref{sec:intro}: the value of \methodname{} does not come from the categorical formatting of any single prompt, but from inducing the vocabulary once and reusing it across procedures and traces.

This diagnostic isolates formatting from content in one small setting. It is not used as a main claim. The result suggests that, once the same trace evidence is supplied, the categorical form is not always separable from the content at this sample size. We therefore frame \methodname{}'s categorical structure as a reusable interface across procedures, not as a claim that code formatting always beats token-matched prose in one prompt.

\clearpage

\end{document}